\documentclass[trackchanges]{aastex701}
\usepackage{placeins}
\usepackage{float}
\usepackage{caption}
\usepackage[normalem]{ulem}

\usepackage{xcolor}
\definecolor{forestgreen(web)}{rgb}{0.13, 0.55, 0.13}

\usepackage{amsmath}
\usepackage{wrapfig}
\begin{document}

\title{Interaction of Fast Magnetoacoustic Wave with the Localized Coronal Null and Generation of the Energetic Alfv\'en Wave Packet}

\author[]{Akash Bairagi}
\affiliation{Department of Physics, Indian Institute of Technology (BHU), Varanasi-221005, India.}
\email{akashbairagi.rs.phy23@itbhu.ac.in}

\author[orcid=0000-0002-1641-1539]{Abhishekh K. Srivastava}
\affiliation{Department of Physics, Indian Institute of Technology (BHU), Varanasi-221005, India. Email:- asrivastava.app@itbhu.ac.in}
\email[show]{asrivastava.app@itbhu.ac.in}

\author[orcid=0009-0000-3578-8270]{Sripan~Mondal}
\affiliation{Department of Physics, Indian Institute of Technology (BHU), Varanasi-221005, India.}
\email{sripan24884@gmail.com}

\author[orcid=0000-0001-5015-5762]{T.V.~Zaqarashvili}
\affiliation{Institute of Physics University of Graz, A-8010 Graz, Austria.}
\affiliation{Department of Astronomy and Astrophysics at Space Research Center, Ilia State University, Kakutsa Cholokashvili Ave 3/5, 0179 Tbilisi, Georgia.}
\affiliation{Evgeni Kharadze Georgian National Astrophysical Observatory, Mount Kanobili, 0301 Abastumani, Georgia.}
\email{teimuraz.zaqarashvili@uni-graz.at}

\author[orcid=0000-0003-2073-002X]{Astrid~Veronig}
\affiliation{Institute of Physics University of Graz, A-8010 Graz, Austria.}
\email{astrid.veronig@uni-graz.at}

\author[orcid=0000-0002-6793-601X]{P.~Bourdin}
\affiliation{Institute of Physics University of Graz, A-8010 Graz, Austria.}
\email{philippe.bourdin@uni-graz.at}

\author[orcid=0000-0002-9514-6402]{Ding Yuan}
\affiliation{Key Laboratory of Solar Activity and Space Weather, School of Aerospace,  Harbin Institute of Technology, Shenzhen, Guangdong, 518055, China. Email:- yuanding@hit.edu.cn}
\email[show]{yuanding@hit.edu.cn}

\author[orcid=0000-0002-2106-9168]{Ryun-Young Kwon}
\affiliation{Korea Astronomy and Space Science Institute, Daejeon 34055, Republic of Korea. Email:- rkwon@kasi.re.kr}
\email[show]{rkwon@kasi.re.kr}
%\author[orcid=0000-0003-3621-6690]{T.V.~Zaqarashvili}
%\affiliation{Institute of Physics University of Graz, A-8010 Graz, Austria | Institute of Physics University of Graz, A-8010 Graz, Austria}
%Solarwind1\email[show]{}

%\collaboration{all}{The Terra Mater collaboration}

\begin{abstract}

%The Sun's atmosphere possesses different kinds of MHD wave modes and their mutual conversion is possible under appropriate plasma and magnetic-field conditions. The fast and slow mode waves are ubiquitously observed in the Sun's corona but the Alfv\'en modes are very difficult to observe due to their incompressible nature. Owing to their properties, the Alfv\'en mode is believed to be one of the energy sources for the solar corona and this mode can even be generated in the vicinity of the equi-partition layer ($c_s\approx c_A$).
%\AB{\sout{through the interaction of a fast mode wave with a magnetic null}}. 
In the present paper, we have performed 2.5D resistive magnetohydrodynamic simulations of the interaction of a fast magnetoacoustic wave with a localized coronal magnetic null point. As a result, an Alfv\'en wave packet is generated by the mode conversion when a %\AB{\sout{EUV wave-like perturbations, {\it akin} of the}} 
fast magnetoacoustic perturbation interacts with the null point. The field-aligned plasma flows are also generated due to the non-linear effects. When the fast mode wavefront interacts with the null, some parts of this wavefront get refracted around it, while some other part is trapped at the null region. Subsequently, the velocity fluctuation out of the plane and in-phase magnetic field fluctuations have evolved and propagated with the local Alfv\'en speed along the separatrixes at one side of the coronal null region. The resulting disturbance behaves as an incompressible and energetic Alfv\'en wave packet. A secondary fast magnetoacoustic wave is also produced and propagates. In the synthetic SDO/AIA observations, no intensity fluctuations are evident in the region where the Alfv\'en wave packet propagates, while the fast magnetoacoustic wave fronts are clearly evident. Our results suggest that given the appropriate physical conditions at the null, when the fast mode wave is incident, Alfv\'en packets can be excited due to the mode conversion, further carrying substantial momentum and energy flux in the solar corona. 

\end{abstract}



\section{Introduction} \label{1}
From many decades a multitude of observational studies have been reported, suggesting that different kinds of magnetohydrodynamic (MHD) wave and oscillation are ubiquitous in the solar atmosphere \citep[e.g.,][and references cited therein] {2002A&A...387L..13D,2002SoPh..209...61D,2005LRSP....2....3N,2006A&A...449L..35K,2007Sci...318.1577O,2007Sci...318.1574D,2007A&A...474..627Z,2011SSRv..158..289G,2023LRSP...20....1J}. According to theoretical studies, in the uniform medium the fast, slow magnetoacoustic waves and Alfv\'en waves are distinct in nature \citep[e.g.,][]{1994ApJ...435..482P,2010adma.book.....G}, but their mutual correlation is important in the complex inhomogeneous solar atmosphere \citep[e.g.,][]{1991GeoRL..18.1951W}. Through imaging, it is difficult to detect the Alfv\'en waves, but changes in the Doppler shifts allow observation of torsional Alfv\'en wave velocity fluctuations \citep[e.g.,][]{2003A&A...399L..15Z}.

If we observationally manifest the presence of fast magnetoacoustic waves into the solar corona, large-scale coronal EUV waves have been observed by the Extreme-ulraviolet Imaging Telescope (EIT) onboard SOHO \citep[e.g.,][]{1997SoPh..175..571M,1998GeoRL..25.2465T}. They are detected as the bright moving fronts in EUV and soft X-ray images, with speeds mostly between $200-400~\mathrm{km\ s^{-1}}$, but events with $1000~\mathrm{km\ s^{-1}}$ have also been observed \citep[e.g.,][]{2011A&A...532A.151W,2013ApJ...776...58N,2014SoPh..289.4563M}. They are generally interpreted as large-amplitude or shocked fast magnetoacoustic waves that propagate through the corona, as a response to the impulsive energy release and mass motion during flare/CME events \citep[e.g.,][]{2015LRSP...12....3W,2017SoPh..292....7L}. The compression of the plasma at the coronal base and transition region, due to the pressure pulse exerted by the passing waves, causes the enhanced emission front observed in EUV and soft X-ray images. Because fast magnetoacoustic waves can propagate in all directions through the corona, they may appear in EUV images as an expanding three-dimensional dome \citep[e.g.,][]{2010ApJ...716L..57V}, while their outer-coronal counterpart is also observed in the coronagraphs as a propagating white-light disturbance across coronal structures and sometime observed as the halo front of CMEs \citep[e.g.,][]{2013ApJ...766...55K,2014ApJ...794..148K,2015ApJ...799L..29K}. When the coronal wave is shocked, it also disturbs the denser chromosphere, producing the so-called Moreton waves observed in the $H_\alpha$ spectral line \citep[e.g.,][]{1960PASP...72..357M,1968SoPh....4...30U}. Recently, the quasi-periodic fast wave trains were also seen propagating outward from the flaring regions in the solar corona \citep[e.g.,][]{2018ApJ...860...54O,2026arXiv260313456W}. There are possibilities that such fast magnetoacoustic perturbations can interact with the inhomogeneities of the coronal magnetic fields (e.g., nulls, QSLs etc) to further lead the reconnection generated plasma dynamics and heating \citep{2025ApJ...984...36S}.

In particular, a magnetic null is a singular point in the solar atmosphere, where the magnetic induction becomes zero ($|\vec B|\approx 0$). Both potential and non-potential field extrapolations of the photospheric magnetic field provide us with the existence of numerous magnetic nulls in the solar corona, and their properties have been analyzed by \citet{2001A&A...367..339B}. The distribution of null points is different in the different layers of the Sun, and their distribution depends on the complexity of the magnetic topology \citep[e.g.,][]{McLaughlin_2010}. Using potential field extrapolation from MDI magnetograms, approximately $1.7\times10^{-3}~\mathrm{Mm^{-2}}$ null points have been calculated by \citet{2004SoPh..225...21C}. Many dynamical processes have been found to be centered around these magnetic null points, such as magnetic reconnection through the collapse of the null by an external EUV wave like velocity perturbation \citep[e.g.,][]{2025ApJ...984...36S}. \citet{2011A&A...533A..18M} showed that initial heating or cooling of the null points after the nano-flares may lead to the generation of entropy modes, which can be observed in the corona. Another model by \citet{2025ApJ...989..222M} has shown the collapse of the null point by boundary motion and the propagation of fast magnetoacoustic waves through the generation of sausage oscillations of the current sheet. Thus, waves and nulls may dynamically interact with each other, and MHD waves near the magnetic null points can be associated with a variety of physical processes in the coronal plasma.

The interaction of MHD waves with a magnetic null can generate different physical processes, such as mode conversion, mode coupling, etc. When a fast magnetoacoustic wave approaches a magnetic null due to the spatial Alfv\'en-speed gradient, mode conversion between fast, slow, and Alfv\'enic perturbations may occur in the vicinity of the $\beta=1$ layer \citep[e.g.,][]{2006A&A...452.1053Z,McLaughlin_2010}, such that:
\begin{equation}
    \beta=\frac{2}{\gamma}\frac{c_s^2}{c_A^2}
\end{equation}
Here, $\gamma$ represents the adiabatic index with value $5/3$, while $c_s$, $c_A$ are the sound and Alfv\'en speed, respectively. At $\beta=1$, $c_s$ is nearly equal to $c_A$ in the equipartition layer. A series of studies have been performed on MHD waves near 2D and 3D magnetic null points with $\beta=0$ regime \citep[e.g.,][]{2004A&A...420.1129M,2005A&A...435..313M,2006A&A...452..603M,2008SoPh..251..563M}. Those papers mainly focus on the accumulation of linear MHD waves near the null region and distinct properties of the different wave modes. In continuation, through the stratified atmosphere, the generation of a high frequency slow mode wave from the fast magnetoacoustic wave at the null in $\beta\not=0$ regime is shown in the numerical model given by \citet{2017A&A...602A..43S}. Recently, \citet{2024NatCo..15.2667K} has provided observational evidence of the fast to slow mode conversion through a coronal 3D null using data from SDO/AIA. Also, the analytical description by \citet{2011ASInC...2..221C} has suggested that mode conversion from fast to Alfv\'en mode in magnetic regions of the solar atmosphere is possible at and beyond the fast wave reflection height, where $\omega=c_Ak$. Though, at the null point, Alfv\'en waves cannot be crossed, but nonlinear Alfv\'en perturbations can generate longitudinal and transverse velocity perturbations due to the ponderomotive force \citep[e.g.,][]{2013A&A...555A..86T}.

As we know, whatever the source of the waves (e.g., photospheric drivers, {\it in-situ} generation, mode conversion, etc), these MHD candidates are one of the major energy sources, thereby transporting energy in the atmosphere of the sun \citep[e.g.,][]{2015RSPTA.37340261A,2021JGRA..12629097S}. The energetics of fast mode waves and the energy distribution to slow mode near the coronal null under a stratified atmosphere have been modeled by \citet{2017ApJ...837...94T}. The fast and slow magnetoacoustic waves may easily dissipate in the lower corona, while Alfv\'en waves carry energy further to the higher corona and may facilitate heating \citep[e.g.,][]{1974SoPh...39..129W}. Due to the inhomogeneous magnetic topology {\it via} resonant absorption, phase mixing, nonlinear wave-wave interactions, the dissipation of Alfv\'en waves may occur in the upper solar corona to sustain the high temperature overcoming thermal conductive and radiative losses \citep[e.g.,][]{2001PhPl....8.2371G,2002AdSpR..30..471D,2002PhRvE..66b6401Z}. When the amplitude of Alfv\'en waves becomes nonlinear, they drive magneto-sonic waves \citep[e.g.,][]{1997SoPh..175...93N,2006A&A...456L..13Z} and dissipate faster in the solar corona and coronal holes. The attenuation of Alfv\'en waves also depends on the straight and curved magnetic field geometries, where the dissipation rate of Alfv\'en waves may be more dominant on the curved magnetic field lines compared to their straighter portion \citep[e.g.,][]{2007A&A...469.1117G}.

In the present work, through the interaction with the null region, we have concentrated on comprehending the mode conversion process from fast to Alfv\'en wave packets \citep[e.g.,][]{2002PhRvE..66b6401Z}. The goal of this paper is to understand how the magnetic null contributes to the generation of Alfv\'en wave packet. In addition, we address the energy flux of the Alfv\'en wave packet at different points on the separatrix and the cause behind the dissipation of the Alfv\'en wave packet near the null point on the curved parts of the magnetic field. The initial perturbation coming towards the model coronal null is similar to the fast magnetoacoustic wave. A portion of it is trapped within the deformed null, which further facilitates the mode conversion and generation of the Alfv\'en wave packet propagating in the large scale solar corona away from the null region. We organize our paper in the following way: a detailed description of the model and non-ideal MHD equations is given in the Numerical Setup and Methods (Sect.~\ref{2}), all the analysis is covered in the Result section (Sect.~\ref{3}) following the Discussion and Conclusion in the Sect.~\ref{4}. 

\section{Numerical Setup and Methods} \label{2}
We have used the open source 'Message Passing Interface-Adaptive Mesh Refinement-Versatile Advection Code (MPI-AMRVAC)' to perform our numerical experiment \citep[e.g.,][]{2012JCoPh.231..718K, 2014ApJS..214....4P, 2018ApJS..234...30X, 2023A&A...673A..66K, 2024ApJ...968..123Z}. Throughout the simulation, we have included fully ionized hydrogen plasma to mimic the lower corona of the Sun. In our model, we simulate the interaction of fast magnetoacoustic wave-like perturbations with the magnetic singularities to investigate the physical processes occurring near the magnetic null.

The spatial dimensions of our simulation domain in the $x$- and $y$- directions are, respectively, 700 Mm and 200 Mm ($x$= [-600, 100 Mm]; $y$=[0 Mm, 200 Mm]). The duration of the simulation time is $1025~\mathrm{s}$ with a temporal cadence of $5.15~\mathrm{s}$. The grid spacings at the coarse level in the $x$ and $y$ directions are $3.125~\mathrm{Mm}$ and $1.563~\mathrm{Mm}$, respectively. The effective maximum resolution is 3584$\times$2048 grid points after the fourth order refinement with the grid-spacings being, approximately $195~\mathrm{km}$ and $98~\mathrm{km}$ in the $x$ and $y$ direction, respectively.

To study the physical dynamics of the system, we have solved the non-ideal basic MHD equations numerically. For performing temporal integrations, we used two step methods, and to find flux at the cell surface, we have followed the Harten-Lax vanLeer (HLL) scheme \citep{1997JCoPh.135..260H}. The utilized MHD equations are given as:
\begin{equation}
    \frac{\partial\rho}{\partial{t}}+\nabla\cdot(\rho\vec{v})=0
\end{equation}
\begin{equation}
    \frac{\partial\vec{B}}{\partial{t}}+\vec\nabla\cdot(\vec{v}\vec{B}-\vec{B}\vec{v})+\vec\nabla\times(\eta\vec{J})=0
\end{equation}
\begin{equation}
    \frac{\partial}{\partial{t}}(\rho\vec{v})+\nabla\cdot[\rho\vec{v}\vec{v}+({p}+\vec{B}^2/2)\vec{I}-\frac{\vec{B}\vec{B}}{4\pi}]=0
\end{equation}
\begin{equation}
    \begin{split}
    \frac{\partial{e}}{\partial{t}}+\vec\nabla\cdot[{e}\vec{v}+({p}+\vec{B}^2/2)\vec{v}-\frac{\vec{B}\vec{B}}{4\pi}\cdot\vec{v}]=\eta\vec{J}^2-\vec{B}\cdot\vec\nabla\times(\eta\vec{J})+\vec\nabla\cdot\vec{q}+{L}_{rad}
    \end{split}
\end{equation}
where,
\begin{equation}
    {e}=\frac{p}{\gamma-1}+\frac{\rho{v}^2}{2}+\frac{{B}^2}{8\pi}
\end{equation}
\begin{equation}
    \vec\nabla\cdot\vec{B}=0,\ \vec{J}=\vec\nabla\times\vec{B}.
\end{equation}
%\begin{equation}
%    \vec{J}=\vec\nabla\times\vec{B}.
%\end{equation}
Here, $\rho$, $p$, $\vec{B}$, $\vec{v}$, $\eta$ and $\vec{J}$ are plasma mass density, plasma pressure, magnetic field, velocity, magnetic diffusivity and current density, respectively. The heat flux vector ($\vec q$) is $\kappa_{||}\vec \nabla_{||} T$, where $\kappa_{||}$ is thermal conduction tensor along the magnetic field. In addition, radiative cooling ($L_{rad}$) is taken into account for our simulation. As usual, $e$ is the total energy density comprised of internal, kinetic, and magnetic energy densities, respectively. For the hydrogen plasma system, we have considered the adiabatic exponent $\gamma=5/3$. The abundance ratio of hydrogen and helium for our model is 10:1, for which the corresponding density $\rho=1.4m_pn_H$ with $m_p$, $n_H$ being the proton mass and the number density, respectively. The initial uniform temperature and density of our system are set as \begin{math} 10^6\ {\rm K} \end{math} and \begin{math} 2.34\times10^{-15}\ {\rm g\ cm^{-3}} \end{math}. Gravity is not taken into account for our localized coronal dynamics. The normalization values for number density, length, time, temperature, mass density, pressure, magnetic field, velocity, and current density are, respectively, \begin{math} {n^*}=10^9\ {\rm cm^{-3}} \end{math}, \begin{math}{L^*}=10^9\ {\rm cm} \end{math}, \begin{math} {t^*}=85.87\ {\rm s} \end{math}, \begin{math} {T^*}=10^6\ {\rm K} \end{math}, \begin{math} {\rho^*}=1.4n^*m_p=2.34\times10^{-15}\ {\rm g\ cm^{-3}} \end{math}, \begin{math} {P^*}=2.3n^*k_BT^*=0.3174\ {\rm dyne\ cm^{-2}} \end{math}, \begin{math} {B^*}=\sqrt{4\pi P^*}=2\ {\rm G} \end{math}, \begin{math} {V^*}=B^*/{\sqrt{4\pi \rho ^*}}= 116\  {\rm km\ s^{-1}} \end{math} and \begin{math} {J^*}=B^* c/\sqrt{4\pi L^*}= 4.77\  {\rm statA\ cm^{-2}} \end{math}. 

In addition, we have included thermal conduction \begin{math} \kappa_{||}=10^{-6}T^{\frac{5}{2}} \end{math}$~\mathrm{erg\ cm^{-1}\ s^{-1}\ K^{-1}}$ \citep{1962pfig.book.....S} parallel to the magnetic field, which helps to distribute thermal energy along the field, while perpendicular conduction is ignored. We adopted uniform magnetic diffusivity throughout the simulation domain with a normalized value of \begin{math} \eta=2\times10^{-4} \end{math}, which corresponds to the physical value of \begin{math} \eta=2.33\times10^{12}\ {\rm cm^2\ s^{-1}} \end{math} \citep{2024ApJ...963..139M}. We do not consider any current-dependent enhancement of the resistivity. Therefore, the initial uniform magnetic diffusivity is applicable for the entire duration of the simulation. Radiative cooling helps to balance the energy of the coronal abundances. For our model, we used the `Dere' radiative loss function \citep{2009A&A...498..915D}, for which,
\begin{equation}
    {L_{rad}} \propto {\int n_e n_H dV}.
\end{equation}

To mimic the dynamics of the large scale corona, we used continuous boundary condition for the top, right, and left boundaries, such that the normal derivatives of all the physical variables are zero at the boundary, and second order extrapolation of magnetic field has been used across the boundary. The pressure and density are kept fixed to their respective initial conditions and all velocity components are kept zero at the bottom boundary. In addition, $B_x$ and $B_z$ are symmetric at the bottom boundary and $B_y$ kept fixed with the initial value. No background and localized heating has been included in our simulation. In addition, the initial current-free bipolar 2.5-dimensional magnetic field topology \citep[see, e.g.,][]{2020ApJ...891...52S,2025ApJ...989..222M} is given below,

\begin{equation}
    %\begin{split}
    {B_x(x,y)}=B_{0}+ \frac{F}{\pi}\left[ \frac{x-x_0}{(x-x_0)^2+(y-y_0)^2}\\
    -\frac{x+x_0}{(x+x_0)^2+(y-y_0)^2}\right]
    %\end{split}
\end{equation}
\begin{equation}
    %\begin{split}
    {B_y(x,y)}=\frac{F(y-y_0)}{\pi}\left[\frac{(x+x_0)^2}{(x-x_0)^2+(y-y_0)^2}\\
    -\frac{(x-x_0)^2}{(x+x_0)^2+(y-y_0)^2}\right]
    %\end{split}
\end{equation}
\begin{equation}
    {B_z}=0.5{B^*}
\end{equation}

%%%%%%%%%%%Fig:1%%%%%%%%%%%%%%%%%%%%%%%%%%%%%%%%%%
\begin{figure*}

\mbox{
\hspace{0.0 cm}
\includegraphics[height=8.0 cm,trim={0 0 0 0},clip]{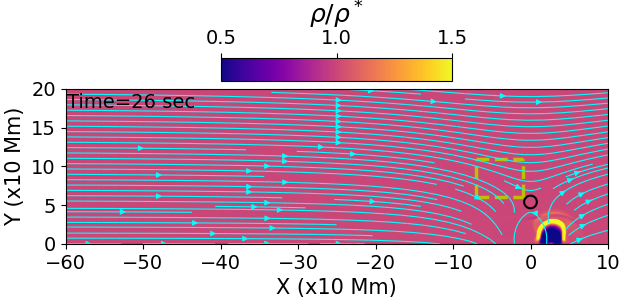}
}

\caption{The density map at $t=26~\mathrm{s}$ is shown with bipolar magnetic field topology over plotted as cyan streamlines. The yellow dashed box is the region of interest for our further analysis and the black contour is a $\beta=1$ region. The fast magnetoacoustic wave-like perturbations is evident in the bottom-right part moving towards the null region. An animation of real time duration $10~\mathrm{s}$ showing the entire dynamics in density from $0$ to $1025~\mathrm{s}$ is available in the online version.}
\label{Fig:1}
\end{figure*}
%%%%%%%%%%%%%%%%%%%%%%%%%%%%%%%%%%%%%%%%%%%%%%%%%%
%%%%%%%%%%%%%FIG:2%%%%%%%%%%%%%%%%%%%%%%%%%%%%%%%
\begin{figure*}
\mbox{
\hspace{-1.4 cm}
\includegraphics[height=21.0 cm,trim={0 0 0 0},clip]{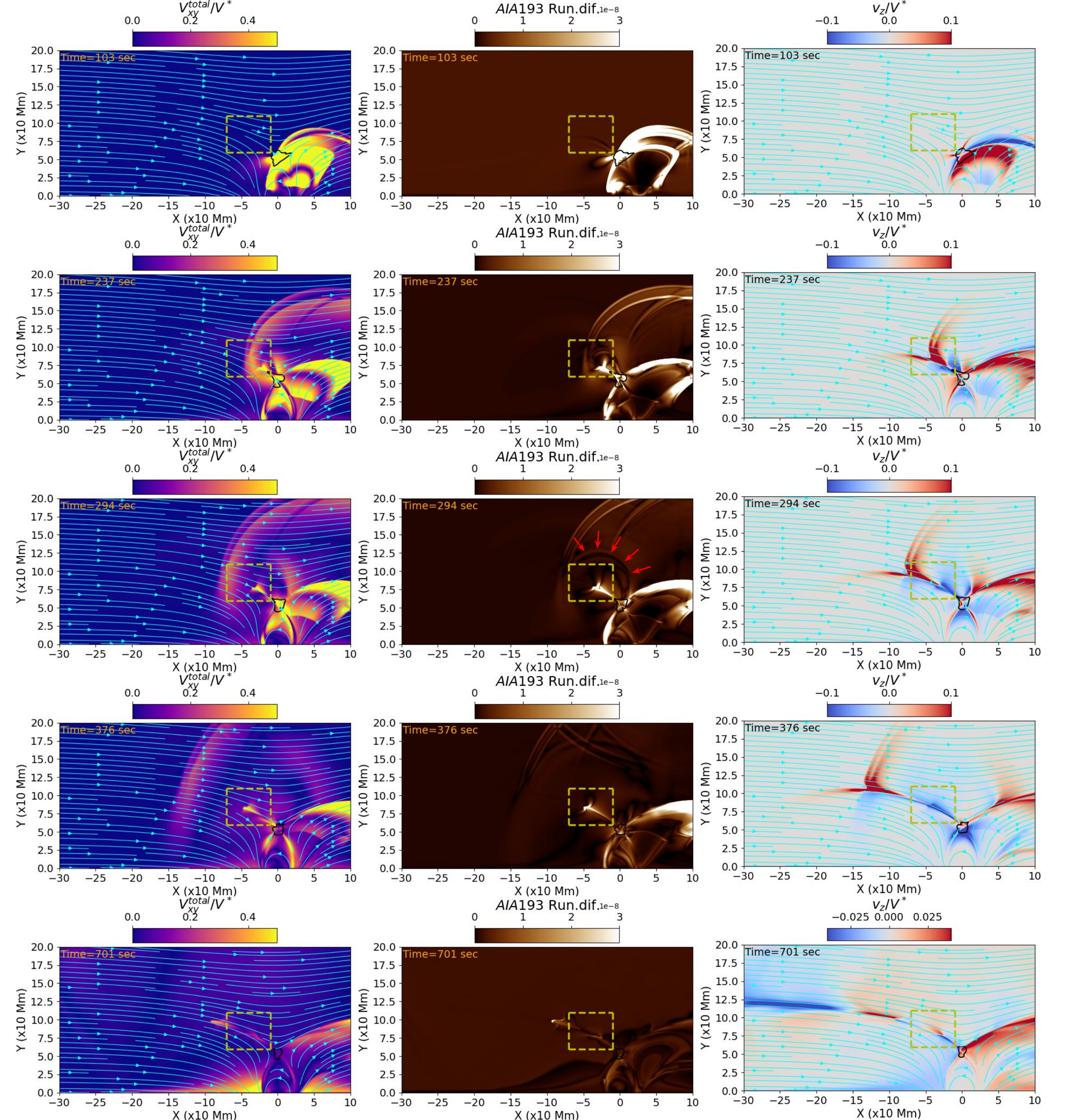}
}  
\caption{First, second and third columns are $v_{xy}^{total}$ (planar velocity), AIA 193 synthesized Intensity, and $v_z$ (out-of-plane velocity) map at $t=103~\mathrm{s},\ 237~\mathrm{s},\ 294~\mathrm{s},\ 376~\mathrm{s}$ and $701~\mathrm{s}$, respectively. At $t=103~\mathrm{s}$, the velocity pulse is incident on the null region (black contour) and further deformed it, which is visible in $v_{xy}^{total}$, synthesized AIA 193 running difference and $v_z$ maps. In the $v_z$ map at $t=237~\mathrm{s}$ and $294~\mathrm{s}$, an Alfv\'en wave packet is also seen to be generated through the $\beta=1$ layer, whereas $v_{xy}^{total}$ and the AIA 193 intensity signal demonstrate the propagation of refracted and newly generated (red arrow) fast magnetoacoustic wave fronts. The independent and distinct propagation of a fast wavefront and Alfv\'en wave packet is seen in the map at $t=376~\mathrm{s}$ and $t=701~\mathrm{s}$. The field-aligned plasma flow is also evident in $v_{xy}^{total}$ and synthesized AIA 193 running difference images surging behind the Alfv\'enic perturbations along the left side separatrix field lines due to non-linear effects. An animation of $v_{xy}^{total}$, AIA 193 running difference and $v_z$ from $0$ to $1025~\mathrm{s}$ are available in the online version. The real time duration of the animation is $10~\mathrm{s}$.}
\label{Fig:2}
\end{figure*}
%%%%%%%%%%%%%%%%%%%%%%%%%%%%%%%%%%%%%%%%%%%%%%%%%%%%%%%

%%%%%%%%%%%%%%%%FIG:3%%%%%%%%%%%%%%%%%%%%%%%%%%%%%%%
\begin{figure*}
\mbox{
\hspace{-1.4 cm}
\includegraphics[height=21.0 cm,trim={0 0 0 0},clip]{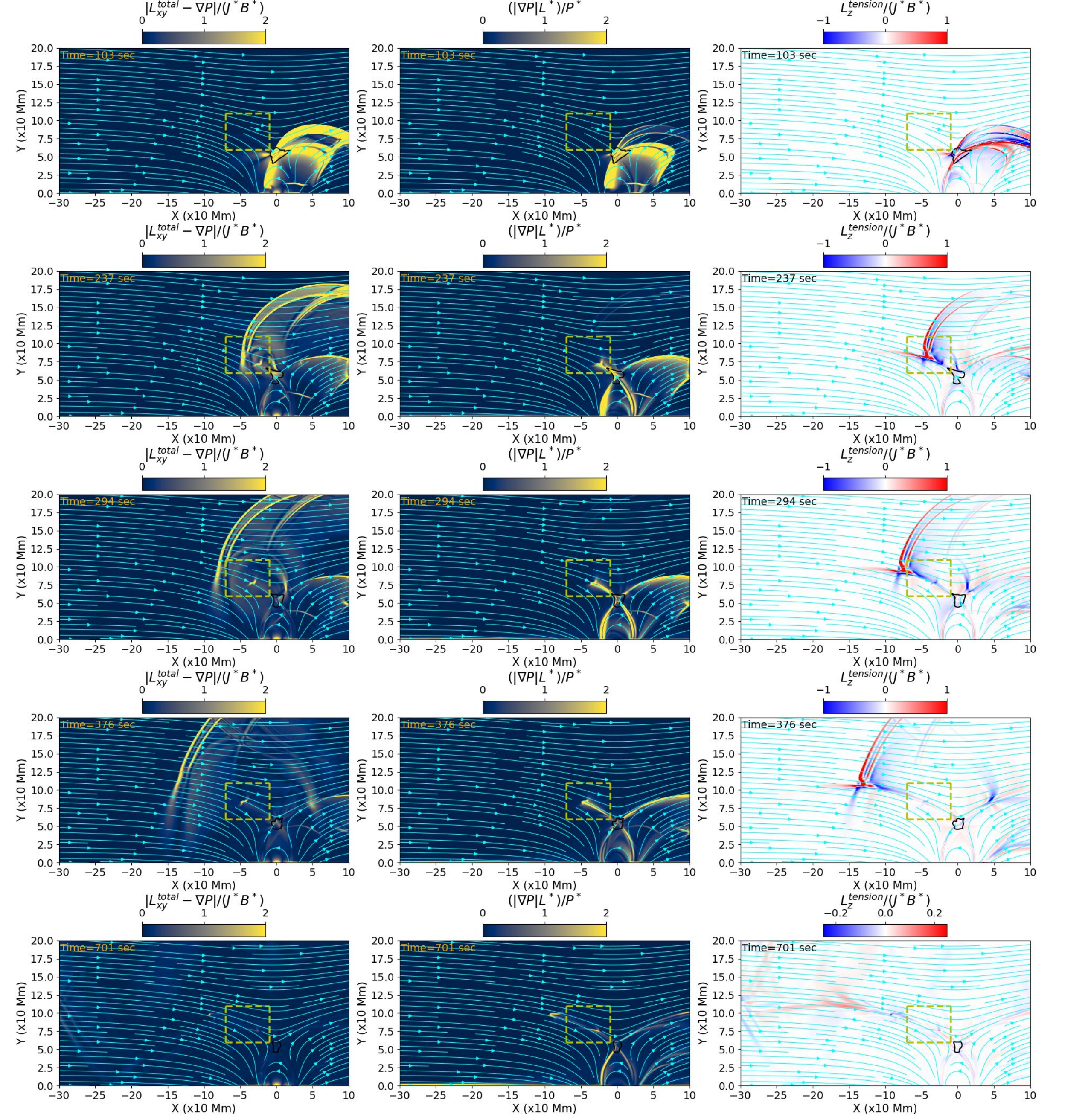}
}  
\caption{First column is the map of total planar Lorentz force ($L_{xy}^{total}$) plus the plasma pressure gradient force ($-\nabla P$) at $t=103~\mathrm{s},\ 237~\mathrm{s},\ 294~\mathrm{s},\ 376~\mathrm{s}$ and $701~\mathrm{s}$, respectively. The signal in the planar Lorentz force has been observed for the fast wavefront propagation, but no detectable signal is present in conjunction with the generation of $v_z$ fluctuations. Also, the magnitude of plasma pressure gradient force maps in the middle column shows the effect very near to the null point for the generation of plasma flow just behind the Alfv\'enic fluctuations. The last columns represents the evolution of the magnetic tension ($L_z^{tension}$) part associated with the Alfv\'en wave packet. The evolution of all the force components from $0$ to $1025~\mathrm{s}$ are available online as an animation with real time duration $10~\mathrm{s}$.}
\label{Fig:3}
\end{figure*}
%%%%%%%%%%%%%%%%%%%%%%%%%%%%%%%%%%%%%%%%%%%%%%%%%%%%%

%%%%%%%%%%%%%%%%FIG:4%%%%%%%%%%%%%%%%%%%%%%%%%%%%%%%%%
\begin{figure*}
\mbox{
\hspace{-0.2 cm}
\includegraphics[height=23.2 cm,trim={0 0 0 0},clip]{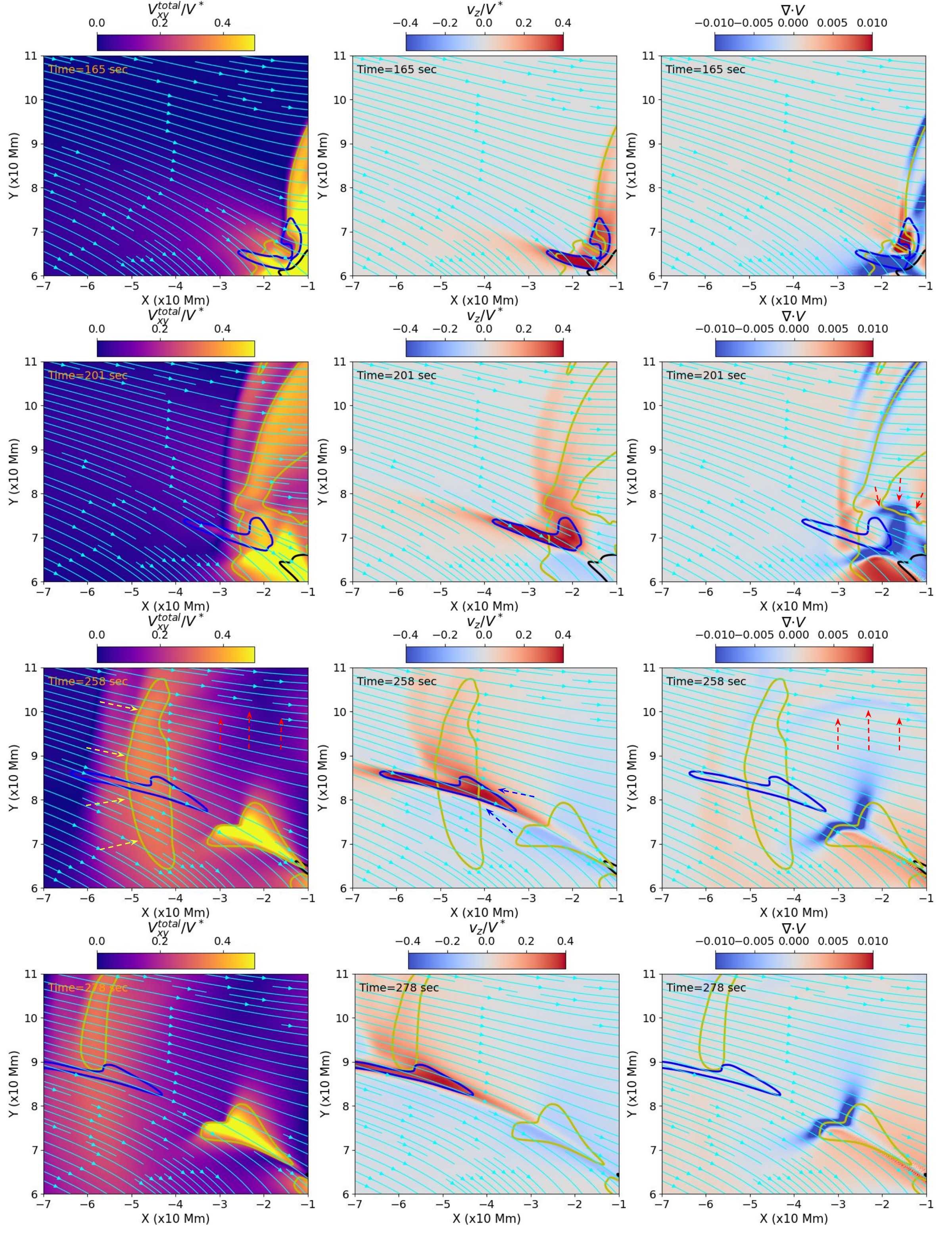}
}  
\caption{From left to right: $v_{xy}^{total}$, $v_z$ and compressibility ($\vec\nabla\cdot\vec v$) on the x-y plane for four time steps. The yellow and blue contour represent the fast magnetoacoustic and Alfv\'en part, respectively to distinguish $v_{xy}^{total}$ and $v_z$ components of the velocity. The associated velocity amplitude range of the contour has been taken $30~\mathrm{km\ s^{-1}}$ to $35~\mathrm{km\ s^{-1}}$. The FOV of all the panel shown here is equivalent to the yellow dotted box shown in Figure~\ref{Fig:3} with dimension [X,Y]=[(-70\ Mm to -10\ Mm), (60\ Mm to 110\ Mm)]. An animation of the above variables from $0$ to $1025~\mathrm{s}$ are available in the online version with real time duration $10~\mathrm{s}$.}
\label{Fig:4}
\end{figure*}
%%%%%%%%%%%%%%%%%%%%%%%%%%%%%%%%%%%%%%%%%%%%%%%%%%%%%%%%%

Here, $F=2\times10^{12}~\mathrm{G\ cm}$, is the flux strength and $B_0=12~\mathrm{G}$ the uniform horizontal magnetic field along the $x$ direction. The two poles are located at $x=5~\mathrm{Mm}$ and $x=-5~\mathrm{Mm}$ (i.e, $x_0=5~\mathrm{Mm}$) in a depth of $20~\mathrm{Mm}$ (i.e, $y_0=-20~\mathrm{Mm}$) from the bottom boundary of the simulation box. In addition, the invariant guide field strength ($B_z$) is initially set to $1~\mathrm{G}$ throughout the simulation domain. 

In the model solar corona, initial $v_y$ pulse further generates large-scale fast waves.
%during solar flares, filament eruptions, and CMEs. Such waves are first discovered via imaging observations in EUV wavebands and therefore termed as EUV waves widely. 
To mimic such fast wave perturbation, a spatial Gaussian velocity pulse has been imposed in the $y$-direction at initial time, such that \citep{2024ApJ...963..139M},
\begin{equation}
    {v_y(x,y,t=0~\mathrm{s})}=v_{0}\exp\left[-\frac{(x-x_{s})^2}{w_x^2}-\frac{(y-y_{s})^2}{w_y^2}\right]
\end{equation}
Here, $(x_s,y_s)=(25, 0)~\mathrm{Mm}$ is the location of the implemented velocity pulse and $v_0=30V^*$ is the velocity amplitude at $t=0~\mathrm{s}$ initially. The pulse is implemented at $t=0~\mathrm{s}$, and is not Gaussian in time. %Instead, it is Gaussian in $x$ and $y$ at $t=0~\mathrm{s}$. 
It is a delta-like time dependence, i.e., non-zero only at the initial time. Although the initial velocity amplitude is large at the epicenter, but when it reaches near the null, the amplitude becomes $60~\mathrm{km\ s^{-1}}$, which is in agreement with observations \citep[e.g.,][]{2012ApJ...752L..23S,2023A&A...675A.129M}. We used a Gaussian velocity pulse that mimics the generation of fast magnetoacoustic waves with a propagation speed around $600~\mathrm{km\ s^{-1}}$ moving in the ambient solar corona. Our motivation is to examine what the physical processes that undergo near the magnetic null when a fast magnetoacoustic perturbations related to the wave are incident on it. The amplitude $3480~\mathrm{km\ s^{-1}}$ is the initial number taken in the numerical experiment that assumes a gigantic blast at the epicenter of the flare-like energy release. We assume a virtual source region where energetic transient activity might source and generate these perturbations. We admit that the initially taken number for the amplitude in the numerical experiment is large locally in the corona, however, it decays very quickly to reach a reasonable value while propagating in the corona (see Appendix also). In principle, we are interested only in the evolution of the velocity pulse {\it akin} of the fast magnetoacoustic waves reaching 54 Mm away from the source (or epicenter) of the disturbances towards the null region where 
%due to the energy distribution in all three directions and expansion 
its amplitude becomes $60~\mathrm{km\ s^{-1}}$ in the vicinity of the null region, which is comparable with the observed magnetoacoustic perturbations in the solar corona \citep[e.g.,][]{2023A&A...675A.129M}. 
%At $t = 0~\mathrm{s}$, the Alfv\'en speed near the epicenter of the velocity perturbations is approximately $700~\mathrm{km\ s^{-1}}$, therefore, the initial amplitude is about $5$ times large that it. However, the physical nature of the velocity driver is such that it quickly decays in amplitude to become sub-Alfv\'enic in the corona, When it is reaching near the null patch its amplitude is of the order of $0.085 c_{A}$, which is a reasonable value. {\it We have included a detailed description in the revised manuscript.
We have analyzed the profile of the velocity driver prior interaction with the magnetic null and added the details of the analysis in the Appendix section. The velocity amplitude of the driver becomes 0.7$c_A$ ($60~\mathrm{km\ s^{-1}}$) at the time of interaction with the null patch, which is a reasonable value of the wave amplitude for the solar corona. It should be noted that the local Alfv\'en speed ($c_A$) is also decreasing, and the $V$/$c_{A}$ values vary spatially while the fast wave is reaching to the null region in the solar corona (refer to the Appendix and its Figure).

We used a single Gaussian spatially dependent velocity pulse at $t = 0~\mathrm{s}$ on one leg of the bipolar magnetic topology (see Figure~\ref{Fig:1}). This pulse evolves spatially as a fast-mode wave and is focused towards the magnetic null due to the Alfv\'en speed gradient. 
Also, the speed of the fast magnetoacoustic wave front depends on the typical coronal plasma conditions and magnetic field value. The spatial widths of the pulse in the $x$ and $y$ directions are $w_x=8~\mathrm{Mm}$ and $w_y=5~\mathrm{Mm}$, respectively. This perturbation gives the response to density and thus perturbing intensity subsequently. Also, from the initial stage a perturbation in the normal component of the velocity ($v_z$) has been generated due to the change of the $B_z$ components which then generates a Lorentz force. The fast magnetic-acoustic perturbations
%, {\it akin} of EUV waves in this study, therefore, 
consist of the resultant perturbations in all the velocity components ($v_{xy}^{total}$ and $v_z$) and also in density (thus also in intensity). On the other hand, pure Alfv\'enic perturbations are truly incompressible, resulting in perturbations in the $v_z$ velocity and $B_z$ field. 
%As we know, Alfv\'en, fast and slow modes are distinct in the uniform medium, but the solar atmosphere is inhomogeneous and those modes can interact with the different kind of magnetic topology. So, the mutual correlation between the modes are important in the sense to understand the energy distribution between them due the interaction with the magnetic singularities. 

In Figure~\ref{Fig:1}, we have shown the density (map) with bipolar magnetic field lines at $t=26~\mathrm{s}$ and the temporal evolution animation of the density from $0$ to $1025~\mathrm{s}$ is available online. The initial perturbation in density is visible at one leg of the bipolar structure and is propagating towards the localized magnetic null, represented as the $\beta=1$ black contour. The yellow dashed box is the region where the Alfv\'en and fast modes are later disentangled. We have also used a uniform guide field ($B_z$) as $1~\mathrm{G}$ in the positive $z$-direction throughout the simulation region. At the null region, the magnetic field strength is negligible, but at the $\beta=1$ region $B_x$ and $B_y$ are not zero, instead they have a comparable value to make the plasma pressure equal to the magnetic pressure. The details of the result are presented in Sect.~\ref{3}.%, and we will discuss and analyze this more in the results section.

\section{Results} \label{3}
After checking the stability of our simulation without any perturbations, a Gaussian velocity pulse %{\it akin} to an EUV wave front 
was implemented below the $\beta=1$ layer at one leg of the magnetic topology. The pulse evolves spatially and temporally and generates propagating fast magnetoacoustic waves, undergoes interaction with the null region, and possible mode conversion from fast to Alfv\'en mode occurs through the vicinity of the equi-partition ($c_s=c_A$) layer. To examine the possible mode conversion from fast to Alfv\'en (basically, quasi-Alfv\'en packet due to the nonzero guide field), we discuss the evolution of different plasma and magnetic properties in the following sub-sections.

\subsection{Temporal evolution of planar velocity ($v_{xy}^{total}$), AIA 193 synthesized intensity and normal velocity component ($v_z$)}\label{3.1}
Figure~\ref{2} and corresponding animation shows the temporal evolution of $v_{xy}^{total}$ ($\sqrt{v_x^2+v_y^2}$) (left column), the running difference of the synthesized AIA 193 intensity images (middle column), and the map of $v_z$ (right column). When a velocity pulse similar to a fast wavefront is implemented for its evolution at the bottom boundary of the simulation box, it starts to propagate towards the magnetic null due to the spatial gradient of the Alfv\'en speed. At $t=103~\mathrm{s}$, it shows that some parts of the $v_{xy}^{total}$ component are trapped at the null region while the others are refracted to the left side of the null, which are also visible in the synthesized AIA 193 intensity, and $v_z$ map. 

It should be noted that fast magnetoacoustic wave-like perturbations possess both $v_{xy}^{total}$ and $v_z$. We provide the Gaussian velocity perturbation initially in the $y$-direction, which later evolves as a fast mode front. But when it propagates, it must perturb the $B_z$, through which $v_z$ arose. When it interacts with the null it can not cross the null, however, from the vicinity of the equipartition layer new $v_z$ fluctuations have formed, which then propagate as an Alfvén wave packet. So, initially the $v_z$ perturbations arise due to the effect of the fast mode perturbation. However, later along the separatrix field lines, only the true incompressible, $v_z$ fluctuations related to the Alfv\'en wave packet is formed and seen propagating. The details of this process is illustrated in forthcoming paragraphs below.

As stated above, after the interaction, the null gets deformed and some part of the front bends and propagates independently from the top of the separatrix towards the left side of the null region. However, at $t=237~\mathrm{s}$ and $294~\mathrm{s}$, the $v_z$ maps depict that a fluctuation in $v_z$ has generated from the vicinity of the equi-partition layer ($c_s=c_A$, black contour) and propagated along the magnetic field with Alfv\'enic speed around $730~\mathrm{km\ s^{-1}}$ to $760~\mathrm{km\ s^{-1}}$ (see sub-section~\ref{3.4}). On the other hand, at $t=294~\mathrm{s}$, a collimated non-periodic plasma flow is left behind the fluctuations $v_z$, which is generated due to the planar Lorentz force ($L_{xy}^{total}$) and the plasma pressure gradient force (see Figure~\ref{Fig:3}) in the neighborhood of the $\beta=1$ layer. %This is generated due to the behavior of a nonlinear Alfv\'en wave on a side of the null region by perturbing the $v_z$
%However, from the initial time of the simulation, the longitudinal velocity components have formed as a result of the perturbation of the longitudinal magnetic field. 
So, those generated longitudinal components do not represent the slow mode, and they are the response of the magnetic field perturbation known as an effect of Ponderomotive force \citep{2013A&A...555A..86T}.

The amplitude of the $v_z$ fluctuation in form of the Alfv\'en wave packet is very high near the $\beta=1$ contour, through which another magnetoacoustic front is generated. This front carried some parts of the energy of the Alfv\'en wave packets to the ambient atmosphere in the upward direction, which is represented by red arrows in the running difference AIA 193 images at $294~\mathrm{s}$ and in agreement with the study of \citet{1997SoPh..175...93N}, for the driving of the higher amplitude magneto-sonic waves due to the nonlinear Alfv\'en waves. At $t=376~\mathrm{s}$ and $701~\mathrm{s}$, the $v_z$ maps depict the independent propagation of the Alfv\'en packet along the magnetic field lines, which is not visible on the $v_{xy}^{total}$ and AIA 193 intensity maps. So, our representation of the generation of the Alfv\'en packet from the null due to the external perturbation follows the theoretical argument for the velocity perturbations ($v_z$). While, the fast mode is accompanied with the intensity fluctuations as seen in the synthetic AIA images, as well as in $v_{xy}^{total}$. 
In the next section, we will show the physical significance of the different components of the Lorentz force and the plasma pressure gradient force for the generation of the wave modes. 

\subsection{Lorentz force and plasma pressure force during the generation of different wave modes}\label{3.2}
In magneto-plasma systems, the Lorentz force plays an important role for the generation of the different wave modes. For our 2.5D system, $L_{xy}^{total}$ is the total Lorentz force or the planar force on the X-Y plane and the normal direction (invariant direction) is only associated with the magnetic tension force ($L_z^{tension}$), while $\vec\nabla P$ is the plasma pressure gradient force (see Figure~\ref{Fig:3} and associated animation). So, here
\begin{equation}
    {|L_{xy}^{total}|}=\sqrt{(J_yB_z-J_zB_y)^2+(J_zB_x-J_xB_z)^2},\ {L_z^{tension}}=(J_xB_y-J_yB_x),\ {|\vec\nabla P|}=\sqrt{(\partial P/\partial x)^2+(\partial P/\partial y)^2}
\end{equation}
Figure~\ref{Fig:3} demonstrates that at $t=103~\mathrm{s}$, the magnitude of the total force is ($|L_{xy}^{total}-\vec\nabla P|$), where $\vec\nabla P$ is significant in the null region due to the incidence of the fast magnetoacoustic wave, but the magnetic tension also has some parts within the null point due to the spatial change of $B_x$ and $B_y$ along the magnetic field. Later, outside the deformed null region, $L_z^{tension}$ is responsible for the generation of fluctuations in the $v_z$ component of the velocity. In the vicinity of the $\beta=1$ layer, the mode conversion from fast to Alfv\'en wave takes place, and all force components are separated corresponding to the different modes. At, $t=237~\mathrm{s},\ 294~\mathrm{s},\ 376~\mathrm{s}$ and $701~\mathrm{s}$, the planar component of force and the plasma pressure force only drive the fast mode and plasma flow through the null region, respectively. However, the magnetic tension component drives the Alfv\'en wave packet along the magnetic field lines. So, all the maps represent significant contributions of all the force components for the generation of different MHD wave modes. Next, we will describe the wave dynamics in the surroundings of the deformed null region, where the possible detachment of different wave modes occurred. %in the neighborhood of null region.

\subsection{Separation of different propagating modes from high to low plasma beta}\label{3.3}
In this section, we demonstrate the behavior of $v_{xy}^{total}$, $v_z$ and $\vec\nabla\cdot \vec v$  (see Figure~\ref{Fig:4} and its animation) between the yellow dashed box-region shown in Figure~\ref{Fig:3}. Here, the yellow contour corresponds to the $v_{xy}^{total}$ component and the blue contour represents the $v_z$ component with the velocity amplitude ranging between $30~\mathrm{km\ s^{-1}}$ and $35~\mathrm{km\ s^{-1}}$.

\textbf {CASE-I} ($t=165~\mathrm{s}$): On the map $v_{xy}^{total}$, one can see that some parts are trapped in the null region and others part bends and crosses the null region and propagates independently. In addition, the $v_z$ signal seen to generate from the equi-partition layer and compressibility has a significant value near the equi-partition layer. So, initially, it is very difficult to separate all the signals in a distinguished manner.

\textbf {CASE-II} ($t=201~\mathrm{s},\ 258~\mathrm{s}$ and $278~\mathrm{s}$): Now as time progresses, the $v_{xy}^{total}$ signal propagates from the high beta to low beta medium as a fast wave (yellow dotted arrows). However, the $v_z$ signal propagates just behind it (see blue arrows), whereas the collimated plasma flow stays behind these two candidates. Another magnetoacoustic front has been generated due to the nonlinear $v_z$ near to the $\beta=1$ layer, which is represented by red arrows in the $v_{xy}^{total}$ and $\vec\nabla\cdot \vec v$ maps at $t=201~\mathrm{s}$ and $258~\mathrm{s}$. This is the secondary fast mode wave front generated due to non-linear effects. From here, at later times, the generated Alfv\'en perturbations accelerate along the magnetic field due to the higher Alfv\'en speed, and all modes are separated and distinguishable at some distance from very close to the null region. Also, the independently propagating fast wavefront is dissipating in the medium, but the Alfv\'en wave is propagating to very far-reaching regions in the solar corona. So, we conclude that the mode conversion is happening through the vicinity of the equi-partition layer and that the planar velocity and compressibility are showing a significant signal for the magnetoacoustic mode and plasma flows. On the other hand, the $v_z$ signal only represents the Alfv\'en mode. 

We must note that it is very difficult to separate $v_z$ and $v_{xy}^{total}$ near the null point initially, because they both are generated simultaneously and Alfv\'en packets accelerate further due to the higher Alfv\'en speed in the low plasma beta region. The $v_{xy}^{total}$ arises due to the dominant plasma pressure and magnetic force near the null region, as shown in Figure~\ref{Fig:3}. We know that in the null region the acoustic nature is dominated and this quasi-stationary plasma flow is driven due to the effect of the plasma pressure gradient ($\vec\nabla P$) force and the planar Lorentz force ($L_{xy}^{total}=|\vec J\times \vec B|_{xy}$), which we have depicted in Figure~\ref{Fig:3}. As the flow is generated from the vicinity of the null region, the plasma pressure gradient force is dominated over the Lorentz force and the flow velocity is field aligned along the separatrix on the $xy$-plane. Also, the field aligned collimated plasma flows are essentially not the entropy mode as they are propagating away from the null region, although with a lesser speed. It is also not a slow mode because there are no periodic longitudinal perturbations observed (see Figure~\ref{Fig:4}). In addition, now we will analyze the properties of the evolved Alfv\'en wave packet in sub-section~\ref{3.4}.  

%%%%%%%%%%%%%%%FIG:5%%%%%%%%%%%%%%%%%%%%%%
\begin{figure*}
\mbox{
\hspace{-1.4 cm}
\includegraphics[height=11.2 cm,trim={0 0 0 0},clip]{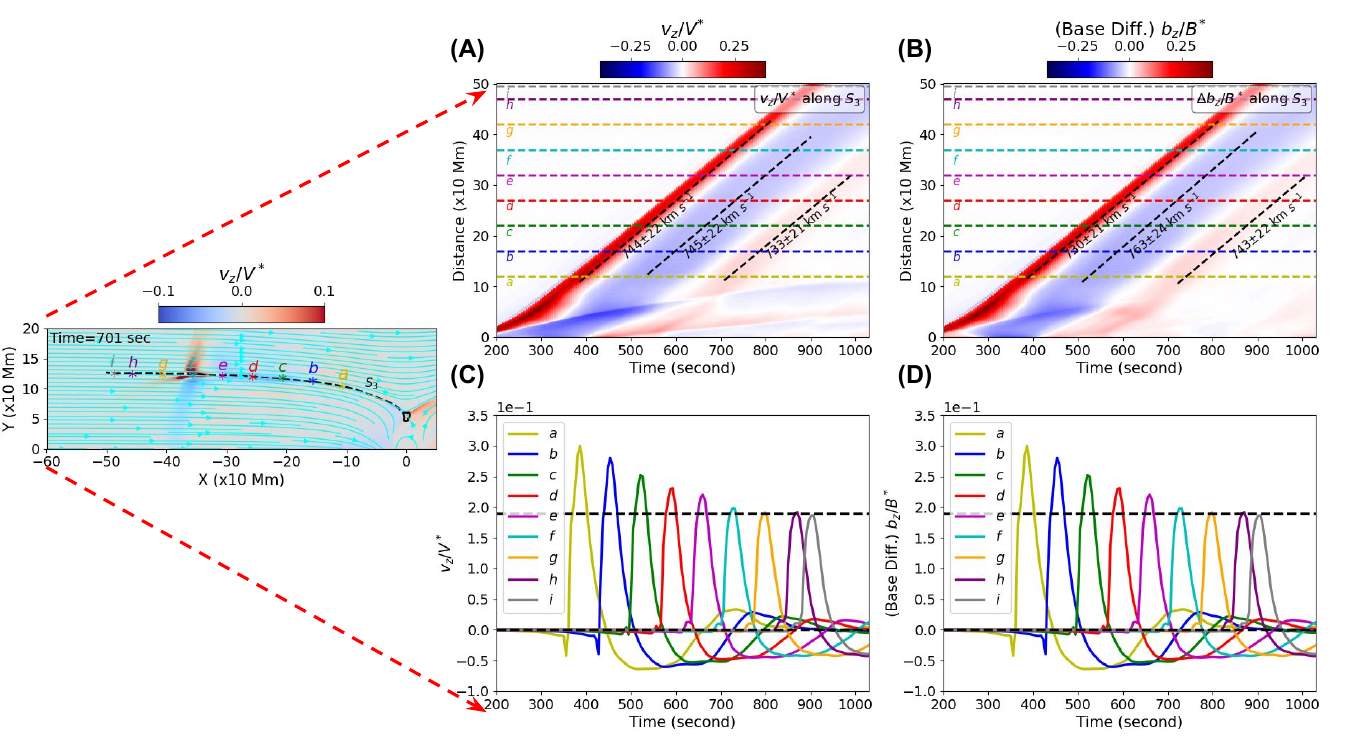}
}  
\caption{The left panel shows the $v_z$ map at $t=701~\mathrm{s}$ with a curved slit $S_3$, starting from right to left along the path of the propagating Alfv\'en wave packet. The independent points a (yellow), b (blue), c (green), d (red), e (magenta), f (cyan), g (orange), h (purple) and i (gray) have been taken on the $S_3$ to understand the properties of $v_z$ at different locations. Panel (A) and (B) are time-distance profiles of $v_z$ and $\Delta b_z=(B_z-0.5)B^*$ along $S_3$ and all the horizontal dotted line representing independent positions on $S_3$ (the line colors kept same as the color of chosen points). Panel (C) and (D) represent the temporal profile of $v_z$ and $\Delta b_z$ at those points and both $v_z$ and $\Delta b_z$ are in the same phase and in agreement with the properties of the Alfv\'en waves.}
\label{Fig:5}
\end{figure*}
%%%%%%%%%%%%%%%%%%%%%%%%%%%%%%%%%%%%%%%%

%%%%%%%%%%%%%%%%%FIG:6%%%%%%%%%%%%%%%%%%%%%%%%%%%%%%%%%%
\begin{figure*}
\mbox{
\hspace{-1.4 cm}
\includegraphics[height=5.4 cm,trim={0 0 0 0},clip]{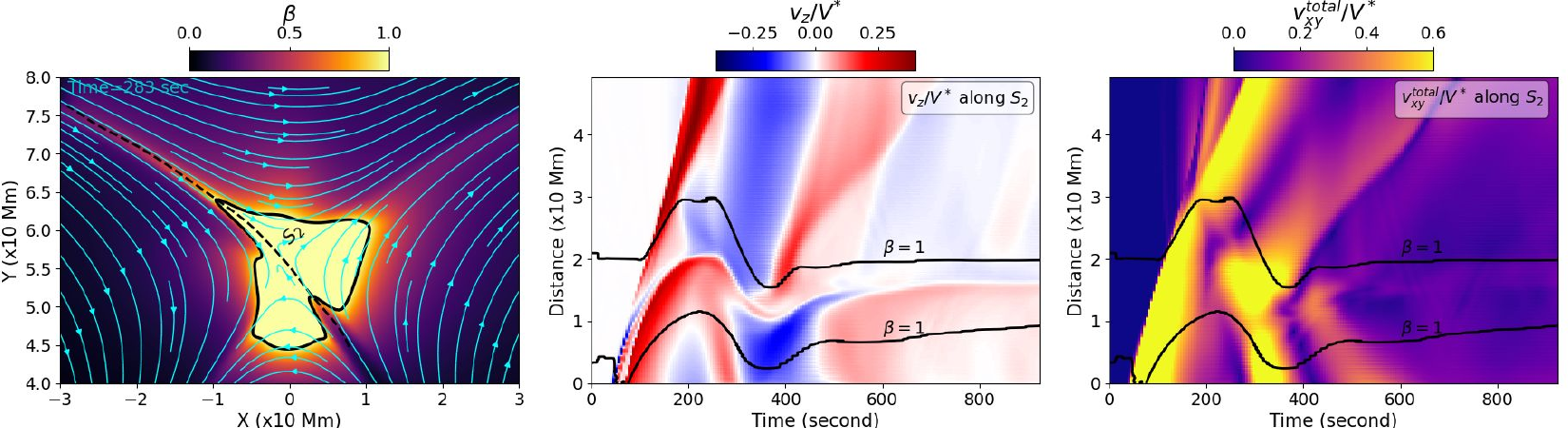}
}  
\caption{Left panel: Zoomed region of the plasma beta map with range 0 to 1 at $t=283~\mathrm{s}$. The black solid contour is $\beta=1$ layer, whereas the black dotted curved slit ($S_2$) is taken from bottom right to top left through the null region. Middle panel: time distance profile with embedded $\beta=1$ layer of $v_z$ along $S_2$. Through the equipartition layer the periodic fluctuations are seen to generate and propagate along the separatrix. Right panel: time distance profile of the planar velocity ($v_{xy}^{total}$) along $S_2$. No periodic fluctuations are observed in $v_{xy}^{total}$ but a flow of plasma is seen surging just behind the $v_z$ signal.}
\label{Fig:6}
\end{figure*}
%%%%%%%%%%%%%%%%%%%%%%%%%%%%%%%%%%%%%%%%%%%%%%%%%%%%%%%

\subsection{Properties of Alfv\'en wave packet along the separatrix}\label{3.4}
In Figure~\ref{Fig:5}, we have analyzed the properties of the Alfv\'en wave packet along the separatrix magnetic field lines extending smoothly into the large-scale corona. To do this analysis, we introduced a curve slit $S_3$ (black dotted line) along the propagation direction of the wave packet, which is visible in the $v_z$ map, with the starting point of the slit taken from the $\beta=1$ layer. As the guide field ($B_z$) is non zero throughout the simulation, the fluctuations in $B_z$ is represented by $\Delta b_z=(B_z-0.5)B^*$. The time distance profile of $v_z$ and $\Delta b_z$ along $S_3$ is in the same phase and shows a uniform periodic propagation. Also, the propagation speed of the crest and trough of the Alfv\'en wave packet in response to $v_z$ and $\Delta b_z$ is $730~\mathrm{km\ s^{-1}}$ to $760~\mathrm{km\ s^{-1}}$, respectively (see Figure~\ref{Fig:5}(A) and (B)). The propagation speed of the packets is Alfv\'enic in nature. Now, different positions like 'a (yellow)', 'b (blue)', 'c (green)', 'd (red)', 'e (magenta)', 'f (cyan)', 'g (orange)', 'h (purple)' and 'i (gray)' on $S_3$ have been considered and these locations are depicted as dotted horizontal lines in the time-distance map to find out the profile of $v_z$ and $\Delta b_z$. In addition, Figure~\ref{Fig:5}(C) and (D) represent the temporal evolution of $v_z$ and $\Delta b_z$ at those fixed locations and at each point periodic perturbations propagate with the same periodicity. The amplitude of $v_z$ is decreasing, at those points that exist on the curve magnetic field and the Alfv\'en packets propagate with uniform amplitude along the horizontal magnetic field without any kind of dissipation. This findings is in agreement with the study by \citet{2007A&A...469.1117G}, that the dissipation of the Alfv\'en wave is more prominent in the curved magnetic tube compared to the linear magnetic tube. In summary, we can say that the generated Alfv\'en wave packet from the null region shows dissipation (60 \% of the initial energy; see sub-section~\ref{3.6}) at the curved magnetic field and propagate transversely with the Alfv\'enic speed to a very far distance without any kind of dissipation along the horizontal magnetic field line. The cause of this dissipation may be due to phase mixing for the Alfv\'en speed gradient across the separatrix of the curved magnetic field line. This study requires a specific set-up and more detailed analyses keeping the scientific aspects of phase-mixing at central place. In the next sub-section we will show that how the mode conversion was initiated through the null region.

\subsection{Importance of magnetic null for the fast to Alfv\'en wave mode conversion}\label{3.5}
Magnetic null points or singularities exist at different layers of the Sun's atmosphere and multiple studies suggested that they can play an important role for different wave mode conversions. In Figure~\ref{Fig:6}, we show how the null region generates Alfv\'en mode from the fast mode incident on it. The first panel (left-most) of Figure~\ref{Fig:6} is the zoomed map of the $\beta$ region with a limit from 0 to 1. The solid black line is the contour $\beta=1$, which deformed after interaction with the fast wave. To understand the temporal behavior of the null, we implemented a curved slit $S_2$ (black dotted line) from the bottom separatrix to the upper separatrix passing through the null intersecting it almost diagonally. The middle panel is the time distance profile of $v_z$ along $S_2$ with a $\beta=1$ contour. The interesting thing is that the area of the contour in the left-panel does not change with time, and it comes back to its original position after 1.5 oscillations. During this oscillation, the periodic fluctuations of $v_z$ (red-blue-red pattern) have generated from the vicinity of the equi-partition layer and propagate along the separatrix as an Alfv\'en wave packet. Within the null region, below $\beta=1$, this velocity component is accompanied by incident fast magnetoacoustic waves (along with the planar velocity component), where it clearly shows the parabolic profile indicating deceleration and trapping partly. Also, the oscillation period of the $\beta=1$ contour is nearly the same as the period of the Alfv\'en wave packet generated and moving ahead above the $\beta=1$ region. Similarly, the third panel is the time-distance profile of the planar velocity ($v_{xy}^{total}$) component, which represents the propagation of the magnetoacoustic mode within the null region as also depicted above. As we can see, during the oscillation of the $\beta=1$ layer, no periodic perturbation has been generated in the planar component of velocity through the $\beta=1$ layer, but a collimated flow of plasma is surging just behind the $v_z$ signal due to the effect of the planar Lorentz force and the plasma pressure gradient force (see Figure~\ref{Fig:3}). So, the $v_z$ signal associated with Alfv\'en mode is completely independent with the $v_{xy}^{total}$, during the propagation from the high beta to the low beta plasma in the neighborhood corona. This means that, due to the presence of a guide field ($B_z$), the magnetic null plays a key role for the generation of Alfv\'en mode from the fast mode due to a mode conversion process in the high beta null region. In continuation, next we will analyze the energetics of the Alfv\'en wave packet.

At this point, it is also important to describe the behavior of the plasma in the null region and surrounding as established in our numerical model. The numerical resistivity is defined as $\eta_{num} \propto c_A\delta x$, where $\delta x$ is the minimum grid size achieved after the final level of AMR refinements. Given the large spatial domain of the Sun's corona and typical Alfv\'en speed (minimum $60~\mathrm{km\ s^{-1}}$ at the null patch and $700~\mathrm{km\ s^{-1}}$ in the nearby corona) based on typical physical conditions, and achieving the best spatial resolution of $\delta x =195~\mathrm{km}$ after four level AMR refinements, we reach the numerical resistivity in the range of $10^{14}~\mathrm{cm^2\ s^{-1}}$ to $10^{15}~\mathrm{cm^2\ s^{-1}}$. The range of such a spatial resolution of a few hundred kilo-meter is typical for modeling the dynamics of the large scale corona \citep{2025A&A...696A.158L}. The physical resistivity is $10^{12}~\mathrm{cm^2\ s^{-1}}$ in the given frame-work, which further demonstrates that the numerical resistivity is dominant in the present case. Given all these conditions at work, the Lundquist number ($S$) of the system is $3.6\times10^{3}$. Such a moderate values of the Lundquist number are also used in modelling the coronal X-point/Current-sheet dynamics, and dissipative Alfv\'enic motions etc in various numerical models \citep[e.g.,][references cited therein]{1995JGR...10023413O,2022A&A...666A..28S}. These physical conditions are able to generate the Alfv\'en wave packets from the null region due to mode conversion and their propagation in the near by region. After the deformation and oscillation of 1.5 cycle, the null patch and surroundings regions seize their oscillatory motions diffusing through the plasma.

%%%%%%%%%%%%%%%%FIG:7%%%%%%%%%%%%%%%%%%%%%%%%%%%%%%%%%%%%%%%%%%%%% 
\begin{figure*}

\mbox{
\hspace{0.5 cm}
\includegraphics[height=12.0 cm,trim={0 0 0 0},clip]{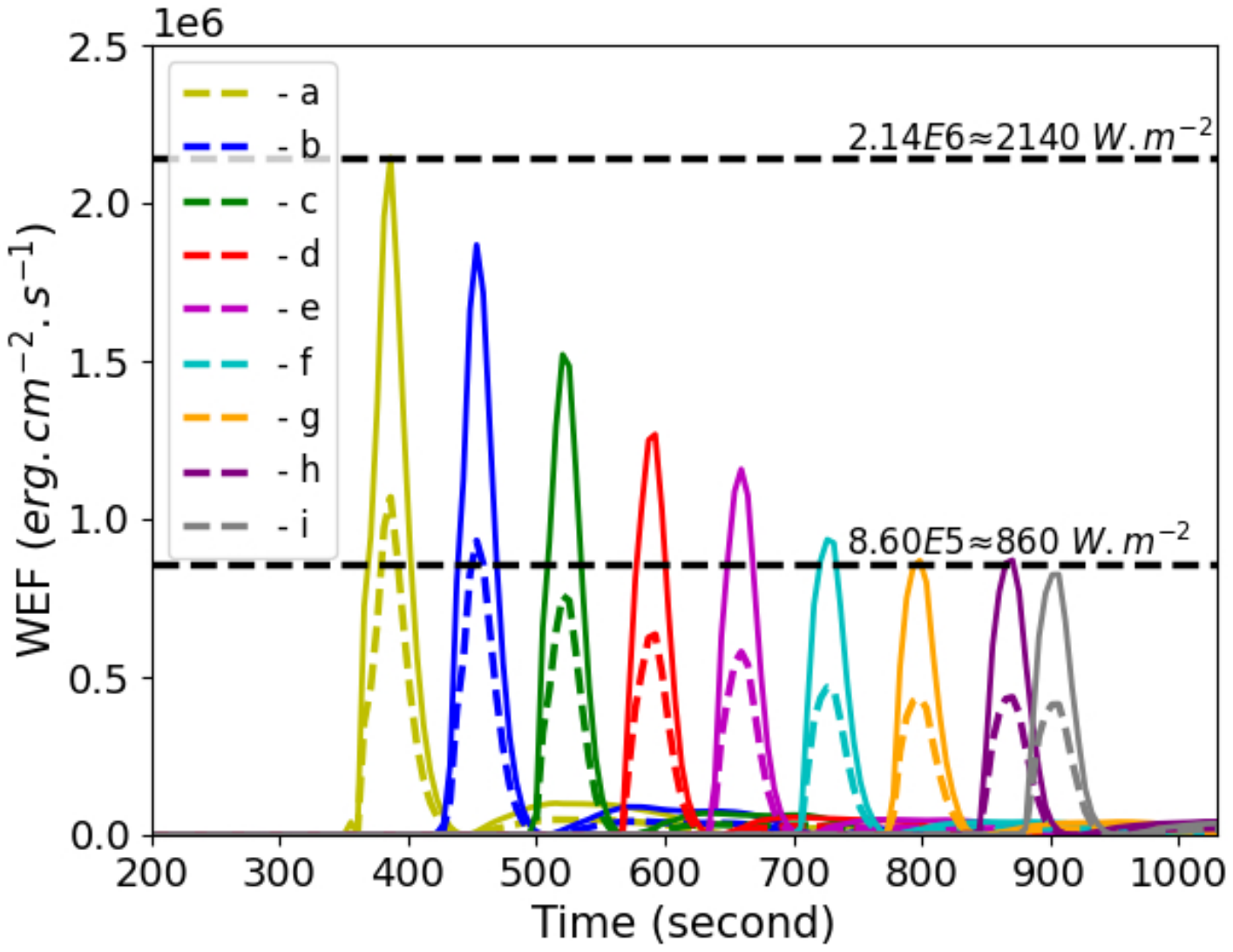}
}

\caption{Temporal evolution of the wave energy flux (WEF) for the Alfv\'en wave packet at the different positions on the slit $S_3$ (see Figure~\ref{Fig:5}). Here, the dotted and solid line respectively correspond to kinetic and total energy (kinetic+magnetic) associated with the WEF. At point 'a' the packet has an energy flux of about $2140~\mathrm{W\ m^{-2}}$, which then decreases gradually at the different points due to the curved magnetic field. Thereafter it reaches a constant value of $860~\mathrm{W\ m^{-2}}$ and propagates up to very far distance along the horizontal field line.}
\label{Fig:7}
\end{figure*}
%%%%%%%%%%%%%%%%%%%%%%%%%%%%%%%%%%%%%%%%%%%%%%%%%%%%%%%

\subsection{Alfv\'en wave energy flux at different locations}\label{3.6}
In order to characterize the energy carried out by the Alfv\'en packet in the medium, we calculate the Wave Energy Flux (WEF) at those fixed locations on the slit $S_3$ (see Figure~\ref{Fig:5}). The WEF is defined as \citep[e.g.,][]{2007Sci...318.1574D,2017NatSR...743147S},
\begin{equation}
    {WEF}=\left[\frac{1}{2}\rho v_z^2+\frac{(\Delta b_z)^2}{8\pi}\right]v_A
\end{equation}
Here, the first and second terms are associated with the kinetic and magnetic part, respectively. Also, $v_z$, $\Delta b_z$ are the perturbation of the $v_z$ and $B_z$ amplitude due to the propagation of the Alfv\'en packets and $\rho$, $v_A$ are the background plasma density and Alfv\'en phase speed (similar to local $c_A$), respectively. When the Alfv\'en packets pass through the separatrix, some parts of the nonlinear Alfv\'en wave energy is transferred to the magnetoacoustic wave, as a result dissipation of some part of the energy flux has occurred on the curved magnetic field. In Figure~\ref{Fig:7}, we show the temporal evolution of the total energy (solid line) and kinetic energy (dotted line) associated with the WEF. At point 'a', the WEF of the packet is approximately $2140~\mathrm{W\ m^{-2}}$ and then the energy starts to decrease due to the curved magnetic field, and further energy remains constant to a value $860~\mathrm{W\ m^{-2}}$ at the locations that exist on the horizontal field lines. The fractional dissipation of the wave energy in the curved part of the magnetic field may be most likely due to the phase-mixing process \citep{2014ApJ...793...43C}, while rest of the energy remain in the propagating wave-packet. This amount of energy suggests that the packet may be energetically significant for coronal energy transport to realize coronal heating, solar wind acceleration, and compensation of radiative losses \citep{1977ARA&A..15..363W}. This means that, the Alfv\'en wave packet generated through mode conversion can carry a sufficient amount of energy to a very far distance.  

\section{Discussions and Conclusion}\label{4}
In this work, we present a comprehensive study on fast magnetoacoustic wave-like perturbations to Alfv\'en mode conversion through a magnetic null point. Under 2.5D magnetic configuration and coronal plasma conditions, an initial velocity perturbation in the $y$ direction has been implemented, which then interacts with the null and undergoes a possible mode conversion to Alfv\'en packet near to the equi-partition layer. The properties of the Alfv\'en wave packet are as follows:

(i) The $v_z$ signal is seen to propagate along the separatrix, while no corresponding signal in intensity and planar velocity ($v_{xy}^{total}$) are evident (see Figure~\ref{Fig:2}). 

(ii) Both $v_z$ and $\Delta b_z$ perturbations associated with the Alfv\'en packet show a high spatial and temporal correlation along the separatrix and are in phase throughout the motion along the magnetic field (see Figures~\ref{Fig:5}(A) and (B)) \citep{2002AdSpR..30..471D}. 

(iii) The propagation speed of the crest and trough that correspond to the Alfv\'en packet is in the range between $730~\mathrm{km\ s^{-1}}$ and $760~\mathrm{km\ s^{-1}}$, which is consistent with an Alfv\'enic interpretation and calculated from the time-distance map of $v_z$ along $S_3$ (see Figure~\ref{Fig:5}).

(iv) Planar projection w.r.t. the direction of the guide-field $B_z$ depicts the Alfv\'enic nature of these evolved transverse wave-packets. 

(v) Finally, these wave-packets contain the sufficient amount of energy to fulfill the radiative losses of the localized solar corona.

Previously, numerous studies have been performed on different mode generation near the null point, such as \citet{2017A&A...602A..43S,2017ApJ...837...94T} shown fast to high frequency slow mode formation around the 2D null point in the stratified model atmosphere. It was also shown that initial heating or cooling of the null points due to the reconnection during nano-flares  may lead to the generation of entropy modes
which can be observed in the corona \citep{2011A&A...533A..18M}.
\citet{2022A&A...660A..21Y} performed a 3D simulation on wave propagation near the magnetic null and provided us with a new mode decomposition technique. Another 2D model gives insight that, under the $\beta=0$ condition, nonlinear Alfv\'en velocity perturbations can generate transverse velocity perturbations around the null due to the effect of the ponderomotive force \citep{2013A&A...555A..86T}.  A series of papers by \citet{2002PhRvE..66b6401Z,2008SoPh..251..251C,2011ASInC...2..221C,2012ApJ...751...31H} has suggested, that Alfv\'en waves can easily be generated through mode conversion from a fast magnetoacoustic wave at and beyond fast wave reflection height in the solar atmosphere. The Alfv\'en waves can be also generated near $\beta \approx 1$ from slow magnetoacoustic waves through nonlinear wave-wave interaction \citep{2006A&A...452.1053Z}. 

For our case, we tried to find out how null point plays an important role for fast magnetoacoustic pulse to Alfv\'en wave-packet mode conversion. The Alfv\'en wave can easily travel from the photosphere to the corona with less dissipation, so this wave acts as an energy source for many magneto-plasma processes, such as the acceleration of the solar wind \citep{1978SoPh...56..305H,2006A&A...456L..13Z}, acceleration of particles \citep{2004ApJ...605L.149V} and many more. In addition, fast magnetoacoustic waves in the form of large-scale EUV waves can be initiated by flare/CME events \citep{2015LRSP...12....3W,2017SoPh..292....7L}, but these waves are not the major energy sources of the corona. During propagation, these waves convert to the other wave modes due to the interaction with the magnetic singularities (e.g., nulls, QSLs, etc). Further, the converted modes are able to carry energy and heat the solar corona under different dissipation mechanisms. However, in the present work, an interesting physical condition became evident uniquely. The fast magnetoacoustic wave-like perturbation in form of a Gaussian pulse mimicked the propagation of an externally driven, EUV wave-like large-scale perturbations \citep{2016ApJ...822..106C,2016SoPh..291.3195C}. When it is incident on a singularity (like a null region with a patch of high plasma beta) within the large scale corona with a guide field, it gets partly trapped, partly refracted, and finally mode conversion takes place to generate an Alfv\'en wave packet. The uniform guide field throughout the simulation only assists the evolved Alfv\'enic fluctuations through the mode conversion. However, 2.5D simulation of external velocity driven reconnection in the coronal current sheet have also been performed by \citet{2024ApJ...963..139M,2025ApJ...979..207M}, which is not our case in the present work. Whereas, \citet{2013A&A...555A..86T} used an ideal 2.5D system in their 2D magnetic null model, to perform the Alfv\'en to fast wave generation from the vicinity of the magnetic null, as a contrary example. 

In future more realistic 3D simulations are required to understand these fluctuations and their inter-conversion in detailed manner. Definitely, the 2.5-D is a simpler set-up compared to full 3-D simulations, where null is a singular patch lying in the X-Y plane instead of the null-ring or any other complex discontinuities that may present in 3D \citep[e.g.,][references cited therein] {2001A&A...367..339B,2003Natur.425..692S,2006AdSpR..37.1269D,2013ApJ...774..154P}. However, it is sufficient to have 2-D null patch and magnetic separatrices in the plane to understand the mode conversion in the current set-up. Another simplification is that a fast wave has a planer projection of the Gaussian envelope and almost circular patches propagating in the given plane, whereas in 3-D the wave would appear as spherical or Gaussian dome-shaped fronts propagating up with different speed \citep[e.g.,][]{2009SSRv..149..153O}. Another point is that visual magnetic field perturbations in form of complex, diverging, and twisted magnetic topologies could be realized in the generated Alfv\'en waves as always in form of toroidal waves \citep[e.g.,][]{2013A&A...555A..86T} rather than the  simple velocity and magnetic field perturbations only out-of-plane as estimated in 2.5-D simulations in the form of true incompressible mode. Moreover, the different values of the energy fluxes will be present in various wave modes considering the presence of the full 3$^{rd}$ dimension \citep[e.g.,][references cited therein]{2022ApJ...924..126S,2022A&A...660A..21Y}. These limitations in the current 2.5-D simulation do not hinder the clear physics of the fast to Alfv\'en wave mode conversion and its finest details obtained in the current model. The full 3-D numerical modeling will be a future objective by implementing various realistic magnetic discontinuities in place.

According to \citet{1977ARA&A..15..363W}, the total energy losses (conduction+radiation) in the quiet region, coronal hole and active region in the corona are, approximately $300~\mathrm{W\ m^{-2}}$, $70~\mathrm{W\ m^{-2}}$ and $10^4~\mathrm{W\ m^{-2}}$. \citet{2009A&A...501L..15B}, has calculated the WEF of the upward propagating Alfv\'en wave from the coronal polar region approximately, $1850~\mathrm{W\ m^{-2}}$, which may heat the corona and accelerate the solar wind. In addition, \citet{2007Sci...318.1574D} estimated that the Alfv\'enic energy flux in the chromosphere level is of the order of $4$ to $7$ $\mathrm{kW\ m^{-2}}$, which is enough to heat the quiet corona and accelerate the solar wind. In our case, we have measured the WEF of the Alfv\'en packet at each location in the direction of it's propagation. Initially, the WEF at region 'a' was $2140~\mathrm{W\ m^{-2}}$, then the energy dissipates on the curve magnetic field, but on the horizontal magnetic field the energy becomes non-dissipative in nature and propagates up to a very far distance with value $860~\mathrm{W\ m^{-2}}$ (see Figure~\ref{Fig:7}). The dissipation of WEF is greater on the curve magnetic field, which is in agreement with the study by \citet{2007A&A...469.1117G}. The initial energy of the Alfv\'en wave packet was $2140~\mathrm{W\ m^{-2}}$, which was more than sufficient to balance the radiative coronal losses and heat it locally. Even after the dissipation of the energy and its reduction to the value of $860~\mathrm{W\ m^{-2}}$, the energy flux is enough to overcome the conductive and radiative energy loss in the corona and can heat the quiet corona and accelerate the solar wind in coronal holes. If, those energy is responsible for the heating, then we need to understand what are the energy dissipation mechanism which is present in the atmosphere?. From many decades phase mixing \citep{1983A&A...117..220H,1997SoPh..173...31I} and resonant absorption \citep{1978ApJ...226..650I} have been believed to be the main energy dissipation mechanism of Alfv\'en waves. Moreover, the turbulent dissipation of the Alfv\'en wave packets, particularly as investigated by \citet{2017NatSR...714820M,2019ApJ...882...50M} may also be a key mechanism for the heating of the solar corona and the acceleration of the solar wind. These literature reported that unidirectional Alfv\'en wave packets, traveling in the same direction, can generate turbulence and dissipate energy, even without the presence of traditional counter-propagating waves, due to non-linear self-interaction across magnetic field inhomogeneities. In our case, we do not consider any of these processes and these are a subject of our future study. %Moreover, we append a note regarding our claim on the signficance of mode-conversion enabled Alfv\'en wave packets on the coronal heating aspect. 
Here we have shown that the fast-mode front undergoes mode conversion in the null region to the Alfv\'en wave packet, which is not a periodic and continuous Alfv\'en wave. The Alfv\'enic fluctuations related to the wave packet are only flowing in a localized spatial domain through the separatrix along the magnetic field lines, so we have calculated the point-wise $'$W$'$ to understand the energy carried out by the Alfv\'en wave packet at different locations momentarily on the separatrix. 
This is the reason why we have shown the point-wise energy distribution of the Alfv\'en wave packet so that its contribution to transporting the packed energy within it to different spatial locations lying on the magnetic separatrix of the localized corona could be estimated. Also, the point-wise $'$W$'$ are in the order of $10^3~\mathrm{W\ m^{-2}}$, and we depict that the total energy of the packet will be in this range to compensate for losses in the localized solar corona. We do not claim here that it will be sufficient to solve the global aspect of energy transport and subsequent heating of the entire corona. However, recurrent Alfv\'en wave packets triggered above the null region due to mode conversion, if dissipated, can produce the sufficient amount of energy to heat the solar corona locally \citep{2017NatSR...714820M,2019ApJ...882...50M}.

From the observation point of view, mode conversion from fast to slow magnetoacoustic wave at a coronal null point has been demonstrated by \cite{2024NatCo..15.2667K}. Alfv\'en waves are ubiquitously present in the solar atmosphere, and they may be generated from mode conversion, but it is still very difficult to detect and visualize Alfv\'en waves through mode conversion in the localized corona due to instrumental limitations. There are numerous nulls present at different heights in the solar atmosphere including the corona, and EUV wave like perturbations are also present in some or other forms sweeping out such random field corona with specific magnetic field topologies. Therefore, such fast to Alfv\'en mode conversion and the {\it in situ} generation of the Alfv\'en waves/packets are highly inevitable processes. Numerically, we can easily distinguish the Alfv\'en signal and their generation should depend on the magnetic singularities and topologically confined regions in the solar corona. Right now we have implemented non periodic velocity perturbations, which evolve in time and interact with the magnetic null results in the generation of a single Alfv\'en wave packet from the vicinity of equi-partition layer. In the future, we will perform the simulation with a periodic velocity perturbation in time, and we expect that multiple Alfv\'en packets will be generated from the null region, which will further carry the continuous momentum and energy from the null region and extend up to a very large distance in the medium. We expect to observe this process and detection of associated dynamics through the upcoming advance coronal spectrometers, like Multi-slit Solar Explorer (MUSE) \citep{2022ApJ...926...52D}. MUSE may be able to detect the existence of Alfv\'enic fluctuations in the vicinity of coronal nulls generated {\it via} this mode conversion process, and capture the associated dynamics \citep{2026ApJ...00...00M}.

%\begin{acknowledgments}
\section*{Acknowledgments}
Authors thank the reviewer for his/her valuable remarks that improved the manuscript. TVZ and FB were supported by Austria-India collaboration project WTZ Indien IN 01/25. AKS, AB, and SM also acknowledge the support of Indo-Austrian Project for this research. AB acknowledges the TA support from IIT (BHU). SM thanks the PMRF fellowship and research grant for supporting his scientific research. D.Y. is supported by the National Natural Science Foundation of China (NSFC: 12473050), the Guangdong Natural Science Funds for Distinguished Young Scholars (2023B1515020049), the Shenzhen Science and Technology Project (JCYJ20240813104805008), and the Specialized Research Fund for the State Key Laboratory of Solar Activity and Space Weather. TVZ was also supported by the Austrian Science Fund (FWF) project PAT7550024 and by Shota Rustaveli National Science Foundation of Georgia (project FR-21-467). R.-Y.K. acknowledges support by basic research funding from the Korea Astronomy and Space Science Institute (KASI; KASI2026185005). 

\appendix
%%%%%%%%%%%%%%%%FIG:8%%%%%%%%%%%%%%%%%%%%%%%%%%%%%%%%%%%%%%%%%%%%% 
\begin{figure*}

\mbox{
\hspace{-1 cm}
\includegraphics[height=12.0 cm,trim={0 0 0 0},clip]{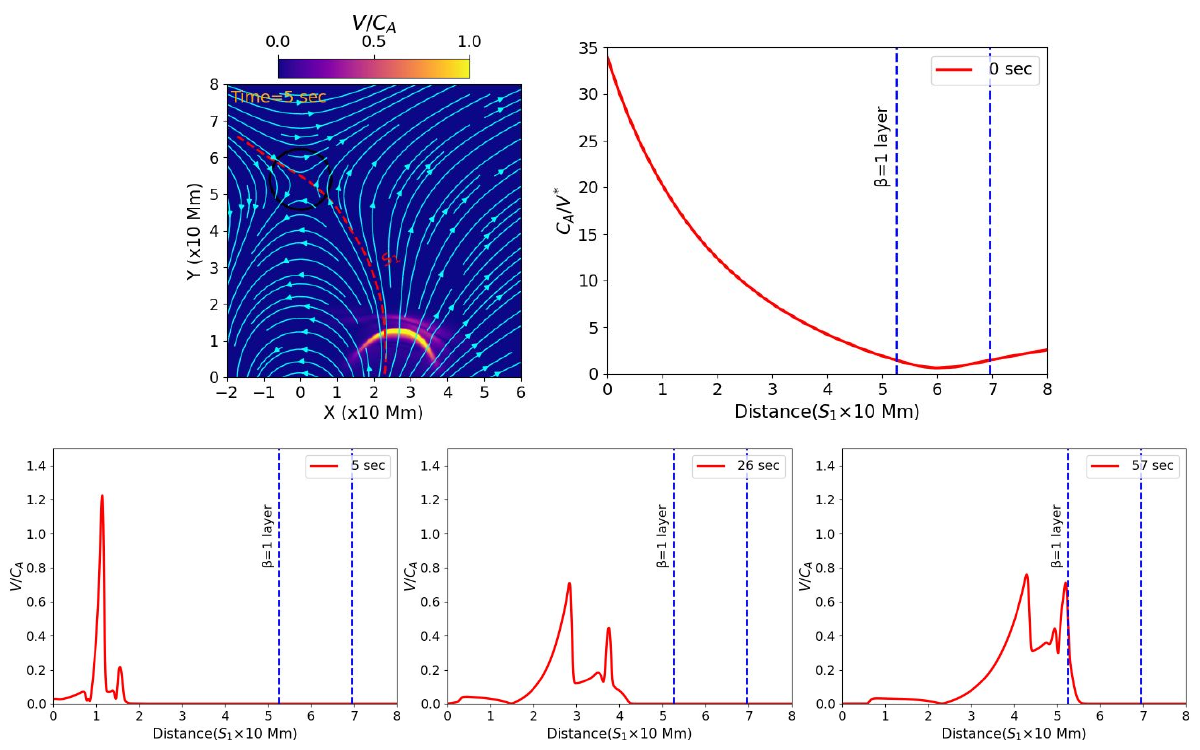}
}

\caption{Top left panel is $v/c_A$ map of the fast magnetoacoustic perturbation at $t = 5~\mathrm{s}$ and dotted red line represents a curved slit chosen from bottom to top through the null region. The FOV of the zoomed region of the numerical domain is, $x=[-20~\mathrm{Mm}, 60~\mathrm{Mm}]$, $y=[0~\mathrm{Mm}, 80~\mathrm{Mm}]$, respectively, and the black circular contour represent the $\beta=1$ layer. Top right panel is the spatial variation of the normalized Alfv\'en speed ($c_A/V^*$) along $S_1$ at $t=0~\mathrm{s}$, and vertical blue dotted line represent the $\beta = 1$ layer on the path of $S_1$. Bottom panel is the spatial characteristics of the ratio of magnetoacoustic perturbation amplitude and local Alfv\'en speed ($v/c_A$) on the slit $S_1$ at $t = 5~\mathrm{s}$, $t = 26~\mathrm{s}$ and $t = 57~\mathrm{s}$, respectively.}
\label{Fig:8}
\end{figure*}
%%%%%%%%%%%%%%%%%%%%%%%%%%%%%%%%%%%%%%%%%%%%%%%%%%%%%%%%%%%%%%%%%%

\section{Evolution of the fast magnetoacoustic perturbations prior interaction with the magnetic null}
In Figure~\ref{Fig:8} we have shown the map of $v/c_A$ (top left) along with a curved slit $S_1$ (red dotted line), started from the one leg of the magnetic topology at the bottom boundary of the numerical domain for tracking the fast-mode perturbation prior to its interaction with the null. Also, we have shown the background profile of the normalized Alfv\'en speed ($c_A/V^*$) along $S_1$ at $t=0~\mathrm{s}$. The Alfv\'en speed profile is not uniform for our model, as it has high value at the bottom layer of the numerical domain due to the higher magnetic field strength. However, as we go towards the region of interest (i.e., null region) and surroundings of the null region, local Alfv\'en speed drops at each instance (e.g., see top-right panel of Figure~\ref{Fig:8} for $t=0~\mathrm{s}$). The $v/c_A$ profile along $S_1$ is showing the amplitude of the fast-mode perturbation normalized with the local Alfv\'en speed. During the evolution of the perturbation towards the magnetic null, the amplitude of the fast-mode front at the time ($57~\mathrm{s}$) of interaction with the $\beta=1$ layer is of the order of $0.7c_A$, which is a reasonable wave amplitude value of $60~\mathrm{km\ s^{-1}}$. 
%So, the perturbation amplitude of the fast-mode front is comparable with the local Alfv\'en speed that has decreasing value at each instances (e.g., see top-right panel of Figure~\ref{Fig:8} for $t=0~\mathrm{s}$) while we move towards the magnetic null. 
Finally, the fast wave undergoes a mode conversion process through the magnetic null region.
%\section{Appendix information}

%\clearpage
\FloatBarrier
\bibliographystyle{aasjournalv7}
\bibliography{sample701}

\begin{thebibliography}{}
\expandafter\ifx\csname natexlab\endcsname\relax\def\natexlab#1{#1}\fi
\providecommand{\url}[1]{\href{#1}{#1}}
\providecommand{\dodoi}[1]{doi:~\href{http://doi.org/#1}{\nolinkurl{#1}}}
\providecommand{\doeprint}[1]{\href{http://ascl.net/#1}{\nolinkurl{http://ascl.net/#1}}}
\providecommand{\doarXiv}[1]{\href{https://arxiv.org/abs/#1}{\nolinkurl{https://arxiv.org/abs/#1}}}

% type= article
\bibitem[{I. {Arregui}(2015){Arregui}}]{2015RSPTA.37340261A}
{Arregui}, I. 2015, \bibinfo{title}{{Wave heating of the solar atmosphere},}
  Philosophical Transactions of the Royal Society of London Series A, 373,
  20140261, \dodoi{10.1098/rsta.2014.0261}

% type= article
\bibitem[{D. {Banerjee} {et~al.}(2009){Banerjee}, {P{\'e}rez-Su{\'a}rez}, \&
  {Doyle}}]{2009A&A...501L..15B}
{Banerjee}, D., {P{\'e}rez-Su{\'a}rez}, D., \& {Doyle}, J.~G. 2009,
  \bibinfo{title}{{Signatures of Alfv{\'e}n waves in the polar coronal holes as
  seen by EIS/Hinode},} \aap, 501, L15, \dodoi{10.1051/0004-6361/200912242}

% type= article
\bibitem[{D.~S. {Brown} \& E.~R. {Priest}(2001){Brown} \&
  {Priest}}]{2001A&A...367..339B}
{Brown}, D.~S., \& {Priest}, E.~R. 2001, \bibinfo{title}{{The topological
  behaviour of 3D null points in the Sun's corona},} \aap, 367, 339,
  \dodoi{10.1051/0004-6361:20010016}

% type= inproceedings
\bibitem[{P.~S. {Cally}(2011){Cally}}]{2011ASInC...2..221C}
{Cally}, P.~S. 2011, \bibinfo{title}{{Alfv{\'e}n waves are easy: mode
  conversion in magnetic regions},} in Astronomical Society of India Conference
  Series, Vol.~2, Astronomical Society of India Conference Series, 221--227

% type= article
\bibitem[{P.~S. {Cally} \& M. {Goossens}(2008){Cally} \&
  {Goossens}}]{2008SoPh..251..251C}
{Cally}, P.~S., \& {Goossens}, M. 2008, \bibinfo{title}{{Three-Dimensional MHD
  Wave Propagation and Conversion to Alfv{\'e}n Waves near the Solar Surface.
  I. Direct Numerical Solution},} \solphys, 251, 251,
  \dodoi{10.1007/s11207-007-9086-3}

% type= article
\bibitem[{R. {Chandra} {et~al.}(2016){Chandra}, {Chen}, {Fulara}, {Srivastava},
  \& {Uddin}}]{2016ApJ...822..106C}
{Chandra}, R., {Chen}, P.~F., {Fulara}, A., {Srivastava}, A.~K., \& {Uddin}, W.
  2016, \bibinfo{title}{{Peculiar Stationary EUV Wave Fronts in the Eruption on
  2011 May 11},} \apj, 822, 106, \dodoi{10.3847/0004-637X/822/2/106}

% type= article
\bibitem[{P.~F. {Chen} {et~al.}(2016){Chen}, {Fang}, {Chandra}, \&
  {Srivastava}}]{2016SoPh..291.3195C}
{Chen}, P.~F., {Fang}, C., {Chandra}, R., \& {Srivastava}, A.~K. 2016,
  \bibinfo{title}{{Can a Fast-Mode EUV Wave Generate a Stationary Front?},}
  \solphys, 291, 3195, \dodoi{10.1007/s11207-016-0920-3}

% type= article
\bibitem[{P. {Chmielewski} {et~al.}(2014){Chmielewski}, {Murawski}, {Musielak},
  \& {Srivastava}}]{2014ApJ...793...43C}
{Chmielewski}, P., {Murawski}, K., {Musielak}, Z.~E., \& {Srivastava}, A.~K.
  2014, \bibinfo{title}{{Numerical Simulations of Impulsively Generated
  Alfv{\'e}n Waves in Solar Magnetic Arcades},} \apj, 793, 43,
  \dodoi{10.1088/0004-637X/793/1/43}

% type= article
\bibitem[{R.~M. {Close} {et~al.}(2004){Close}, {Parnell}, \&
  {Priest}}]{2004SoPh..225...21C}
{Close}, R.~M., {Parnell}, C.~E., \& {Priest}, E.~R. 2004,
  \bibinfo{title}{{Separators in 3D Quiet-Sun Magnetic Fields},} \solphys, 225,
  21, \dodoi{10.1007/s11207-004-3259-0}

% type= article
\bibitem[{I. {De Moortel} {et~al.}(2002{\natexlab{a}}){De Moortel}, {Ireland},
  {Hood}, \& {Walsh}}]{2002A&A...387L..13D}
{De Moortel}, I., {Ireland}, J., {Hood}, A.~W., \& {Walsh}, R.~W.
  2002{\natexlab{a}}, \bibinfo{title}{{The detection of 3 \& 5 min period
  oscillations in coronal loops},} \aap, 387, L13,
  \dodoi{10.1051/0004-6361:20020436}

% type= article
\bibitem[{I. {De Moortel} {et~al.}(2002{\natexlab{b}}){De Moortel}, {Ireland},
  {Walsh}, \& {Hood}}]{2002SoPh..209...61D}
{De Moortel}, I., {Ireland}, J., {Walsh}, R.~W., \& {Hood}, A.~W.
  2002{\natexlab{b}}, \bibinfo{title}{{Longitudinal intensity oscillations in
  coronal loops observed with TRACE I. Overview of Measured Parameters},}
  \solphys, 209, 61, \dodoi{10.1023/A:1020956421063}

% type= article
\bibitem[{B. {De Pontieu} {et~al.}(2007){De Pontieu}, {McIntosh}, {Carlsson},
  {Hansteen}, {Tarbell}, {Schrijver}, {Title}, {Shine}, {Tsuneta}, {Katsukawa},
  {Ichimoto}, {Suematsu}, {Shimizu}, \& {Nagata}}]{2007Sci...318.1574D}
{De Pontieu}, B., {McIntosh}, S.~W., {Carlsson}, M., {et~al.} 2007,
  \bibinfo{title}{{Chromospheric Alfv{\'e}nic Waves Strong Enough to Power the
  Solar Wind},} Science, 318, 1574, \dodoi{10.1126/science.1151747}

% type= article
\bibitem[{B. {De Pontieu} {et~al.}(2022){De Pontieu}, {Testa},
  {Mart{\'\i}nez-Sykora}, {Antolin}, {Karampelas}, {Hansteen}, {Rempel},
  {Cheung}, {Reale}, {Danilovic}, {Pagano}, {Polito}, {De Moortel},
  {N{\'o}brega-Siverio}, {Van Doorsselaere}, {Petralia}, {Asgari-Targhi},
  {Boerner}, {Carlsson}, {Chintzoglou}, {Daw}, {DeLuca}, {Golub}, {Matsumoto},
  {Ugarte-Urra}, {McIntosh}, \& {the MUSE Team}}]{2022ApJ...926...52D}
{De Pontieu}, B., {Testa}, P., {Mart{\'\i}nez-Sykora}, J., {et~al.} 2022,
  \bibinfo{title}{{Probing the Physics of the Solar Atmosphere with the
  Multi-slit Solar Explorer (MUSE). I. Coronal Heating},} \apj, 926, 52,
  \dodoi{10.3847/1538-4357/ac4222}

% type= article
\bibitem[{L. {Del Zanna} \& M. {Velli}(2002){Del Zanna} \&
  {Velli}}]{2002AdSpR..30..471D}
{Del Zanna}, L., \& {Velli}, M. 2002, \bibinfo{title}{{Coronal heating through
  Alfven waves},} Advances in Space Research, 30, 471,
  \dodoi{10.1016/S0273-1177(02)00320-4}

% type= article
\bibitem[{P. {D{\'e}moulin}(2006){D{\'e}moulin}}]{2006AdSpR..37.1269D}
{D{\'e}moulin}, P. 2006, \bibinfo{title}{{Extending the concept of separatrices
  to QSLs for magnetic reconnection},} Advances in Space Research, 37, 1269,
  \dodoi{10.1016/j.asr.2005.03.085}

% type= article
\bibitem[{K.~P. {Dere} {et~al.}(2009){Dere}, {Landi}, {Young}, {Del Zanna},
  {Landini}, \& {Mason}}]{2009A&A...498..915D}
{Dere}, K.~P., {Landi}, E., {Young}, P.~R., {et~al.} 2009,
  \bibinfo{title}{{CHIANTI - an atomic database for emission lines. IX.
  Ionization rates, recombination rates, ionization equilibria for the elements
  hydrogen through zinc and updated atomic data},} \aap, 498, 915,
  \dodoi{10.1051/0004-6361/200911712}

% type= book
\bibitem[{J.~P. {Goedbloed} {et~al.}(2010){Goedbloed}, {Keppens}, \&
  {Poedts}}]{2010adma.book.....G}
{Goedbloed}, J.~P., {Keppens}, R., \& {Poedts}, S. 2010, {Advanced
  Magnetohydrodynamics}

% type= article
\bibitem[{M. {Goossens} \& A. {De Groof}(2001){Goossens} \& {De
  Groof}}]{2001PhPl....8.2371G}
{Goossens}, M., \& {De Groof}, A. 2001, \bibinfo{title}{{Resonant and
  phase-mixed magnetohydrodynamic waves in the solar atmosphere},} Physics of
  Plasmas, 8, 2371, \dodoi{10.1063/1.1343090}

% type= article
\bibitem[{M. {Goossens} {et~al.}(2011){Goossens}, {Erd{\'e}lyi}, \&
  {Ruderman}}]{2011SSRv..158..289G}
{Goossens}, M., {Erd{\'e}lyi}, R., \& {Ruderman}, M.~S. 2011,
  \bibinfo{title}{{Resonant MHD Waves in the Solar Atmosphere},} \ssr, 158,
  289, \dodoi{10.1007/s11214-010-9702-7}

% type= article
\bibitem[{M. {Gruszecki} {et~al.}(2007){Gruszecki}, {Murawski}, {Solanki}, \&
  {Ofman}}]{2007A&A...469.1117G}
{Gruszecki}, M., {Murawski}, K., {Solanki}, S.~K., \& {Ofman}, L. 2007,
  \bibinfo{title}{{Attenuation of Alfv{\'e}n waves in straight and curved
  coronal slabs},} \aap, 469, 1117, \dodoi{10.1051/0004-6361:20066924}

% type= article
\bibitem[{S.~C. {Hansen} \& P.~S. {Cally}(2012){Hansen} \&
  {Cally}}]{2012ApJ...751...31H}
{Hansen}, S.~C., \& {Cally}, P.~S. 2012, \bibinfo{title}{{Benchmarking
  Fast-to-Alfv{\'e}n Mode Conversion in a Cold MHD Plasma. II. How to Get
  Alfv{\'e}n Waves through the Solar Transition Region},} \apj, 751, 31,
  \dodoi{10.1088/0004-637X/751/1/31}

% type= article
\bibitem[{A. {Harten}(1997){Harten}}]{1997JCoPh.135..260H}
{Harten}, A. 1997, \bibinfo{title}{{High Resolution Schemes for Hyperbolic
  Conservation Laws},} Journal of Computational Physics, 135, 260,
  \dodoi{10.1006/jcph.1997.5713}

% type= article
\bibitem[{J. {Heyvaerts} \& E.~R. {Priest}(1983){Heyvaerts} \&
  {Priest}}]{1983A&A...117..220H}
{Heyvaerts}, J., \& {Priest}, E.~R. 1983, \bibinfo{title}{{Coronal heating by
  phase-mixed shear Alfven waves.},} \aap, 117, 220

% type= article
\bibitem[{J.~V. {Hollweg}(1978){Hollweg}}]{1978SoPh...56..305H}
{Hollweg}, J.~V. 1978, \bibinfo{title}{{Alfv{\'e}n waves in the solar
  atmosphere.},} \solphys, 56, 305, \dodoi{10.1007/BF00152474}

% type= article
\bibitem[{J.~A. {Ionson}(1978){Ionson}}]{1978ApJ...226..650I}
{Ionson}, J.~A. 1978, \bibinfo{title}{{Resonant absorption of Alfv{\'e}nic
  surface waves and the heating of solar coronal loops.},} \apj, 226, 650,
  \dodoi{10.1086/156648}

% type= article
\bibitem[{J. {Ireland} \& E.~R. {Priest}(1997){Ireland} \&
  {Priest}}]{1997SoPh..173...31I}
{Ireland}, J., \& {Priest}, E.~R. 1997, \bibinfo{title}{{Phase-mixing in
  Dissipative Alfv{\'e}n Waves},} \solphys, 173, 31,
  \dodoi{10.1023/A:1004903128146}

% type= article
\bibitem[{D.~B. {Jess} {et~al.}(2023){Jess}, {Jafarzadeh}, {Keys},
  {Stangalini}, {Verth}, \& {Grant}}]{2023LRSP...20....1J}
{Jess}, D.~B., {Jafarzadeh}, S., {Keys}, P.~H., {et~al.} 2023,
  \bibinfo{title}{{Waves in the lower solar atmosphere: the dawn of
  next-generation solar telescopes},} Living Reviews in Solar Physics, 20, 1,
  \dodoi{10.1007/s41116-022-00035-6}

% type= article
\bibitem[{R. {Keppens} {et~al.}(2012){Keppens}, {Meliani}, {van Marle},
  {Delmont}, {Vlasis}, \& {van der Holst}}]{2012JCoPh.231..718K}
{Keppens}, R., {Meliani}, Z., {van Marle}, A.~J., {et~al.} 2012,
  \bibinfo{title}{{Parallel, grid-adaptive approaches for relativistic hydro
  and magnetohydrodynamics},} Journal of Computational Physics, 231, 718,
  \dodoi{10.1016/j.jcp.2011.01.020}

% type= article
\bibitem[{R. {Keppens} {et~al.}(2023){Keppens}, {Popescu Braileanu}, {Zhou},
  {Ruan}, {Xia}, {Guo}, {Claes}, \& {Bacchini}}]{2023A&A...673A..66K}
{Keppens}, R., {Popescu Braileanu}, B., {Zhou}, Y., {et~al.} 2023,
  \bibinfo{title}{{MPI-AMRVAC 3.0: Updates to an open-source simulation
  framework},} \aap, 673, A66, \dodoi{10.1051/0004-6361/202245359}

% type= article
\bibitem[{V. {Kukhianidze} {et~al.}(2006){Kukhianidze}, {Zaqarashvili}, \&
  {Khutsishvili}}]{2006A&A...449L..35K}
{Kukhianidze}, V., {Zaqarashvili}, T.~V., \& {Khutsishvili}, E. 2006,
  \bibinfo{title}{{Observation of kink waves in solar spicules},} \aap, 449,
  L35, \dodoi{10.1051/0004-6361:200600018}

% type= article
\bibitem[{P. {Kumar} {et~al.}(2024){Kumar}, {Nakariakov}, {Karpen}, \&
  {Cho}}]{2024NatCo..15.2667K}
{Kumar}, P., {Nakariakov}, V.~M., {Karpen}, J.~T., \& {Cho}, K.-S. 2024,
  \bibinfo{title}{{Direct imaging of magnetohydrodynamic wave mode conversion
  near a 3D null point on the sun},} Nature Communications, 15, 2667,
  \dodoi{10.1038/s41467-024-46736-4}

% type= article
\bibitem[{R.-Y. {Kwon} {et~al.}(2013){Kwon}, {Ofman}, {Olmedo}, {Kramar},
  {Davila}, {Thompson}, \& {Cho}}]{2013ApJ...766...55K}
{Kwon}, R.-Y., {Ofman}, L., {Olmedo}, O., {et~al.} 2013,
  \bibinfo{title}{{STEREO Observations of Fast Magnetosonic Waves in the
  Extended Solar Corona Associated with EIT/EUV Waves},} \apj, 766, 55,
  \dodoi{10.1088/0004-637X/766/1/55}

% type= article
\bibitem[{R.-Y. {Kwon} {et~al.}(2014){Kwon}, {Zhang}, \&
  {Olmedo}}]{2014ApJ...794..148K}
{Kwon}, R.-Y., {Zhang}, J., \& {Olmedo}, O. 2014, \bibinfo{title}{{New Insights
  into the Physical Nature of Coronal Mass Ejections and Associated Shock Waves
  within the Framework of the Three-dimensional Structure},} \apj, 794, 148,
  \dodoi{10.1088/0004-637X/794/2/148}

% type= article
\bibitem[{R.-Y. {Kwon} {et~al.}(2015){Kwon}, {Zhang}, \&
  {Vourlidas}}]{2015ApJ...799L..29K}
{Kwon}, R.-Y., {Zhang}, J., \& {Vourlidas}, A. 2015, \bibinfo{title}{{Are
  Halo-like Solar Coronal Mass Ejections Merely a Matter of Geometric
  Projection Effects?},} \apjl, 799, L29, \dodoi{10.1088/2041-8205/799/2/L29}

% type= article
\bibitem[{V. {Liakh} \& R. {Keppens}(2025){Liakh} \&
  {Keppens}}]{2025A&A...696A.158L}
{Liakh}, V., \& {Keppens}, R. 2025, \bibinfo{title}{{Numerical study of solar
  eruption, extreme-ultraviolet wave propagation, and wave-induced prominence
  dynamics},} \aap, 696, A158, \dodoi{10.1051/0004-6361/202453300}

% type= article
\bibitem[{D.~M. {Long} {et~al.}(2017){Long}, {Bloomfield}, {Chen}, {Downs},
  {Gallagher}, {Kwon}, {Vanninathan}, {Veronig}, {Vourlidas}, {Vr{\v{s}}nak},
  {Warmuth}, \& {{\v{Z}}ic}}]{2017SoPh..292....7L}
{Long}, D.~M., {Bloomfield}, D.~S., {Chen}, P.~F., {et~al.} 2017,
  \bibinfo{title}{{Understanding the Physical Nature of Coronal ``EIT
  Waves''},} \solphys, 292, 7, \dodoi{10.1007/s11207-016-1030-y}

% type= article
\bibitem[{N. {Magyar} {et~al.}(2017){Magyar}, {Van Doorsselaere}, \&
  {Goossens}}]{2017NatSR...714820M}
{Magyar}, N., {Van Doorsselaere}, T., \& {Goossens}, M. 2017,
  \bibinfo{title}{{Generalized phase mixing: Turbulence-like behaviour from
  unidirectionally propagating MHD waves},} Scientific Reports, 7, 14820,
  \dodoi{10.1038/s41598-017-13660-1}

% type= article
\bibitem[{N. {Magyar} {et~al.}(2019){Magyar}, {Van Doorsselaere}, \&
  {Goossens}}]{2019ApJ...882...50M}
{Magyar}, N., {Van Doorsselaere}, T., \& {Goossens}, M. 2019,
  \bibinfo{title}{{Understanding Uniturbulence: Self-cascade of MHD Waves in
  the Presence of Inhomogeneities},} \apj, 882, 50,
  \dodoi{10.3847/1538-4357/ab357c}

% type= article
\bibitem[{G. {Mann} {et~al.}(2023){Mann}, {Warmuth}, \&
  {{\"O}nel}}]{2023A&A...675A.129M}
{Mann}, G., {Warmuth}, A., \& {{\"O}nel}, H. 2023, \bibinfo{title}{{Kinematical
  evolution of large-scale EUV waves in the solar corona},} \aap, 675, A129,
  \dodoi{10.1051/0004-6361/202346378}

% type= article
\bibitem[{J.~A. {McLaughlin} {et~al.}(2008){McLaughlin}, {Ferguson}, \&
  {Hood}}]{2008SoPh..251..563M}
{McLaughlin}, J.~A., {Ferguson}, J.~S.~L., \& {Hood}, A.~W. 2008,
  \bibinfo{title}{{3D MHD Coronal Oscillations about a Magnetic Null Point:
  Application of WKB Theory},} \solphys, 251, 563,
  \dodoi{10.1007/s11207-007-9107-2}

% type= article
\bibitem[{J.~A. {McLaughlin} \& A.~W. {Hood}(2004){McLaughlin} \&
  {Hood}}]{2004A&A...420.1129M}
{McLaughlin}, J.~A., \& {Hood}, A.~W. 2004, \bibinfo{title}{{MHD wave
  propagation in the neighbourhood of a two-dimensional null point},} \aap,
  420, 1129, \dodoi{10.1051/0004-6361:20035900}

% type= article
\bibitem[{J.~A. {McLaughlin} \& A.~W. {Hood}(2005){McLaughlin} \&
  {Hood}}]{2005A&A...435..313M}
{McLaughlin}, J.~A., \& {Hood}, A.~W. 2005, \bibinfo{title}{{MHD wave
  propagation in the neighbourhood of two null points},} \aap, 435, 313,
  \dodoi{10.1051/0004-6361:20042361}

% type= article
\bibitem[{J.~A. {McLaughlin} \& A.~W. {Hood}(2006){McLaughlin} \&
  {Hood}}]{2006A&A...452..603M}
{McLaughlin}, J.~A., \& {Hood}, A.~W. 2006,
  \bibinfo{title}{{Magnetohydrodynamics wave propagation in the neighbourhood
  of two dipoles},} \aap, 452, 603, \dodoi{10.1051/0004-6361:20054575}

% type= article
\bibitem[{J.~A. McLaughlin {et~al.}(2010)McLaughlin, Hood, \&
  De~Moortel}]{McLaughlin_2010}
McLaughlin, J.~A., Hood, A.~W., \& De~Moortel, I. 2010, \bibinfo{title}{Review
  Article: MHD Wave Propagation Near Coronal Null Points of Magnetic Fields,}
  Space Science Reviews, 158, 205–236, \dodoi{10.1007/s11214-010-9654-y}

% type= article
\bibitem[{S. {Mondal} {et~al.}(2025{\natexlab{a}}){Mondal}, {Bairagi}, \&
  {Srivastava}}]{2025ApJ...979..207M}
{Mondal}, S., {Bairagi}, A., \& {Srivastava}, A.~K. 2025{\natexlab{a}},
  \bibinfo{title}{{Dynamics and Energetics of Resistive, Thermally Conductive,
  and Radiative Plasma in Coronal Current Sheets due to Asymmetric External
  Perturbation},} \apj, 979, 207, \dodoi{10.3847/1538-4357/ada1d6}

% type= article
\bibitem[{S. {Mondal} {et~al.}(2026){Mondal}, {Srivastava}, {Bairagi}, {
  Martinez Sykora}, \& {De Pontieu}}]{2026ApJ...00...00M}
{Mondal}, S., {Srivastava}, A., {Bairagi}, A., { Martinez Sykora}, J., \& {De
  Pontieu}, B. e.~a. 2026, \bibinfo{title}{{On observing the Alfv\'en Wave
  packets in the Vicinity of Coronal Nulls using MUSE'},} ApJL, 000, 00,
  \dodoi{in preparation}

% type= article
\bibitem[{S. {Mondal} {et~al.}(2025{\natexlab{b}}){Mondal}, {Srivastava},
  {Pontin}, \& {Priest}}]{2025ApJ...989..222M}
{Mondal}, S., {Srivastava}, A.~K., {Pontin}, D.~I., \& {Priest}, E.~R.
  2025{\natexlab{b}}, \bibinfo{title}{{Generation of Surface Sausage
  Oscillations of a Current Sheet and Propagating Magnetoacoustic Waves by
  Impulsive Reconnection},} \apj, 989, 222, \dodoi{10.3847/1538-4357/adf18e}

% type= article
\bibitem[{S. {Mondal} {et~al.}(2024){Mondal}, {Srivastava}, {Pontin}, {Yuan},
  \& {Priest}}]{2024ApJ...963..139M}
{Mondal}, S., {Srivastava}, A.~K., {Pontin}, D.~I., {Yuan}, D., \& {Priest},
  E.~R. 2024, \bibinfo{title}{{2.5D Magnetohydrodynamic Simulation of the
  Formation and Evolution of Plasmoids in Coronal Current Sheets},} \apj, 963,
  139, \dodoi{10.3847/1538-4357/ad2079}

% type= article
\bibitem[{G.~E. {Moreton} \& H.~E. {Ramsey}(1960){Moreton} \&
  {Ramsey}}]{1960PASP...72..357M}
{Moreton}, G.~E., \& {Ramsey}, H.~E. 1960, \bibinfo{title}{{Recent Observations
  of Dynamical Phenomena Associated with Solar Flares},} \pasp, 72, 357,
  \dodoi{10.1086/127549}

% type= article
\bibitem[{D. {Moses} {et~al.}(1997){Moses}, {Clette}, {Delaboudini{\`e}re},
  {Artzner}, {Bougnet}, {Brunaud}, {Carabetian}, {Gabriel}, {Hochedez},
  {Millier}, {Song}, {Au}, {Dere}, {Howard}, {Kreplin}, {Michels}, {Defise},
  {Jamar}, {Rochus}, {Chauvineau}, {Marioge}, {Catura}, {Lemen}, {Shing},
  {Stern}, {Gurman}, {Neupert}, {Newmark}, {Thompson}, {Maucherat},
  {Portier-Fozzani}, {Berghmans}, {Cugnon}, {Van Dessel}, \&
  {Gabryl}}]{1997SoPh..175..571M}
{Moses}, D., {Clette}, F., {Delaboudini{\`e}re}, J.-P., {et~al.} 1997,
  \bibinfo{title}{{EIT Observations of the Extreme Ultraviolet Sun},} \solphys,
  175, 571, \dodoi{10.1023/A:1004902913117}

% type= article
\bibitem[{N. {Muhr} {et~al.}(2014){Muhr}, {Veronig}, {Kienreich},
  {Vr{\v{s}}nak}, {Temmer}, \& {Bein}}]{2014SoPh..289.4563M}
{Muhr}, N., {Veronig}, A.~M., {Kienreich}, I.~W., {et~al.} 2014,
  \bibinfo{title}{{Statistical Analysis of Large-Scale EUV Waves Observed by
  STEREO/EUVI},} \solphys, 289, 4563, \dodoi{10.1007/s11207-014-0594-7}

% type= article
\bibitem[{K. {Murawski} {et~al.}(2011){Murawski}, {Zaqarashvili}, \&
  {Nakariakov}}]{2011A&A...533A..18M}
{Murawski}, K., {Zaqarashvili}, T.~V., \& {Nakariakov}, V.~M. 2011,
  \bibinfo{title}{{Entropy mode at a magnetic null point as a possible tool for
  indirect observation of nanoflares in the solar corona},} \aap, 533, A18,
  \dodoi{10.1051/0004-6361/201116942}

% type= article
\bibitem[{V.~M. {Nakariakov} {et~al.}(1997){Nakariakov}, {Roberts}, \&
  {Murawski}}]{1997SoPh..175...93N}
{Nakariakov}, V.~M., {Roberts}, B., \& {Murawski}, K. 1997,
  \bibinfo{title}{{Alfv{\'e}n Wave Phase Mixing as a Source of Fast
  Magnetosonic Waves},} \solphys, 175, 93, \dodoi{10.1023/A:1004965725929}

% type= article
\bibitem[{V.~M. {Nakariakov} \& E. {Verwichte}(2005){Nakariakov} \&
  {Verwichte}}]{2005LRSP....2....3N}
{Nakariakov}, V.~M., \& {Verwichte}, E. 2005, \bibinfo{title}{{Coronal Waves
  and Oscillations},} Living Reviews in Solar Physics, 2, 3,
  \dodoi{10.12942/lrsp-2005-3}

% type= article
\bibitem[{N.~V. {Nitta} {et~al.}(2013){Nitta}, {Schrijver}, {Title}, \&
  {Liu}}]{2013ApJ...776...58N}
{Nitta}, N.~V., {Schrijver}, C.~J., {Title}, A.~M., \& {Liu}, W. 2013,
  \bibinfo{title}{{Large-scale Coronal Propagating Fronts in Solar Eruptions as
  Observed by the Atmospheric Imaging Assembly on Board the Solar Dynamics
  Observatory{\textemdash}an Ensemble Study},} \apj, 776, 58,
  \dodoi{10.1088/0004-637X/776/1/58}

% type= article
\bibitem[{L. {Ofman}(2009){Ofman}}]{2009SSRv..149..153O}
{Ofman}, L. 2009, \bibinfo{title}{{Progress, Challenges, and Perspectives of
  the 3D MHD Numerical Modeling of Oscillations in the Solar Corona},} \ssr,
  149, 153, \dodoi{10.1007/s11214-009-9501-1}

% type= article
\bibitem[{L. {Ofman} \& J.~M. {Davila}(1995){Ofman} \&
  {Davila}}]{1995JGR...10023413O}
{Ofman}, L., \& {Davila}, J.~M. 1995, \bibinfo{title}{{Alfv{\'e}n wave heating
  of coronal holes and the relation to the high-speed solar wind},} \jgr, 100,
  23413, \dodoi{10.1029/95JA02222}

% type= article
\bibitem[{L. {Ofman} \& W. {Liu}(2018){Ofman} \& {Liu}}]{2018ApJ...860...54O}
{Ofman}, L., \& {Liu}, W. 2018, \bibinfo{title}{{Quasi-periodic
  Counter-propagating Fast Magnetosonic Wave Trains from Neighboring Flares:
  SDO/AIA Observations and 3D MHD Modeling},} \apj, 860, 54,
  \dodoi{10.3847/1538-4357/aac2e8}

% type= article
\bibitem[{T.~J. {Okamoto} {et~al.}(2007){Okamoto}, {Tsuneta}, {Berger},
  {Ichimoto}, {Katsukawa}, {Lites}, {Nagata}, {Shibata}, {Shimizu}, {Shine},
  {Suematsu}, {Tarbell}, \& {Title}}]{2007Sci...318.1577O}
{Okamoto}, T.~J., {Tsuneta}, S., {Berger}, T.~E., {et~al.} 2007,
  \bibinfo{title}{{Coronal Transverse Magnetohydrodynamic Waves in a Solar
  Prominence},} Science, 318, 1577, \dodoi{10.1126/science.1145447}

% type= article
\bibitem[{D.~I. {Pontin} {et~al.}(2013){Pontin}, {Priest}, \&
  {Galsgaard}}]{2013ApJ...774..154P}
{Pontin}, D.~I., {Priest}, E.~R., \& {Galsgaard}, K. 2013, \bibinfo{title}{{On
  the Nature of Reconnection at a Solar Coronal Null Point above a Separatrix
  Dome},} \apj, 774, 154, \dodoi{10.1088/0004-637X/774/2/154}

% type= article
\bibitem[{L.~J. {Porter} {et~al.}(1994){Porter}, {Klimchuk}, \&
  {Sturrock}}]{1994ApJ...435..482P}
{Porter}, L.~J., {Klimchuk}, J.~A., \& {Sturrock}, P.~A. 1994,
  \bibinfo{title}{{The Possible Role of MHD Waves in Heating the Solar
  Corona},} \apj, 435, 482, \dodoi{10.1086/174830}

% type= article
\bibitem[{O. {Porth} {et~al.}(2014){Porth}, {Xia}, {Hendrix}, {Moschou}, \&
  {Keppens}}]{2014ApJS..214....4P}
{Porth}, O., {Xia}, C., {Hendrix}, T., {Moschou}, S.~P., \& {Keppens}, R. 2014,
  \bibinfo{title}{{MPI-AMRVAC for Solar and Astrophysics},} \apjs, 214, 4,
  \dodoi{10.1088/0067-0049/214/1/4}

% type= article
\bibitem[{S. {Sabri} {et~al.}(2022){Sabri}, {Ebadi}, \&
  {Poedts}}]{2022ApJ...924..126S}
{Sabri}, S., {Ebadi}, H., \& {Poedts}, S. 2022, \bibinfo{title}{{Propagation of
  the Alfv{\'e}n Wave and Induced Perturbations in the Vicinity of a 3D Proper
  Magnetic Null Point},} \apj, 924, 126, \dodoi{10.3847/1538-4357/ac3b5f}

% type= article
\bibitem[{I.~C. {Santamaria} {et~al.}(2017){Santamaria}, {Khomenko},
  {Collados}, \& {de Vicente}}]{2017A&A...602A..43S}
{Santamaria}, I.~C., {Khomenko}, E., {Collados}, M., \& {de Vicente}, A. 2017,
  \bibinfo{title}{{High-frequency waves in the corona due to null points},}
  \aap, 602, A43, \dodoi{10.1051/0004-6361/201629729}

% type= article
\bibitem[{S. {Sen} \& R. {Keppens}(2022){Sen} \&
  {Keppens}}]{2022A&A...666A..28S}
{Sen}, S., \& {Keppens}, R. 2022, \bibinfo{title}{{Thermally enhanced tearing
  in solar current sheets: Explosive reconnection with plasmoid-trapped
  condensations},} \aap, 666, A28, \dodoi{10.1051/0004-6361/202244152}

% type= article
\bibitem[{Y. {Shen} \& Y. {Liu}(2012){Shen} \& {Liu}}]{2012ApJ...752L..23S}
{Shen}, Y., \& {Liu}, Y. 2012, \bibinfo{title}{{Simultaneous Observations of a
  Large-scale Wave Event in the Solar Atmosphere: From Photosphere to Corona},}
  \apjl, 752, L23, \dodoi{10.1088/2041-8205/752/2/L23}

% type= article
\bibitem[{S.~K. {Solanki} {et~al.}(2003){Solanki}, {Lagg}, {Woch}, {Krupp}, \&
  {Collados}}]{2003Natur.425..692S}
{Solanki}, S.~K., {Lagg}, A., {Woch}, J., {Krupp}, N., \& {Collados}, M. 2003,
  \bibinfo{title}{{Three-dimensional magnetic field topology in a region of
  solar coronal heating},} \nat, 425, 692, \dodoi{10.1038/nature02035}

% type= book
\bibitem[{L. {Spitzer}(1962){Spitzer}}]{1962pfig.book.....S}
{Spitzer}, L. 1962, {Physics of Fully Ionized Gases}

% type= article
\bibitem[{A.~K. {Srivastava} {et~al.}(2017){Srivastava}, {Shetye}, {Murawski},
  {Doyle}, {Stangalini}, {Scullion}, {Ray}, {W{\'o}jcik}, \&
  {Dwivedi}}]{2017NatSR...743147S}
{Srivastava}, A.~K., {Shetye}, J., {Murawski}, K., {et~al.} 2017,
  \bibinfo{title}{{High-frequency torsional Alfv{\'e}n waves as an energy
  source for coronal heating},} Scientific Reports, 7, 43147,
  \dodoi{10.1038/srep43147}

% type= article
\bibitem[{A.~K. {Srivastava} {et~al.}(2021){Srivastava}, {Ballester}, {Cally},
  {Carlsson}, {Goossens}, {Jess}, {Khomenko}, {Mathioudakis}, {Murawski}, \&
  {Zaqarashvili}}]{2021JGRA..12629097S}
{Srivastava}, A.~K., {Ballester}, J.~L., {Cally}, P.~S., {et~al.} 2021,
  \bibinfo{title}{{Chromospheric Heating by Magnetohydrodynamic Waves and
  Instabilities},} Journal of Geophysical Research (Space Physics), 126,
  e029097, \dodoi{10.1029/2020JA029097}

% type= article
\bibitem[{A.~K. {Srivastava} {et~al.}(2025){Srivastava}, {Mondal}, {Priest},
  {Mishra}, {Pontin}, {Kwon}, {Yuan}, {Murawski}, \&
  {Asai}}]{2025ApJ...984...36S}
{Srivastava}, A.~K., {Mondal}, S., {Priest}, E.~R., {et~al.} 2025,
  \bibinfo{title}{{Localized Heating and Dynamics of the Solar Corona due to a
  Symbiosis of Waves and Reconnection},} \apj, 984, 36,
  \dodoi{10.3847/1538-4357/adc379}

% type= article
\bibitem[{P. {Syntelis} \& E.~R. {Priest}(2020){Syntelis} \&
  {Priest}}]{2020ApJ...891...52S}
{Syntelis}, P., \& {Priest}, E.~R. 2020, \bibinfo{title}{{A Cancellation
  Nanoflare Model for Solar Chromospheric and Coronal Heating. III. 3D
  Simulations and Atmospheric Response},} \apj, 891, 52,
  \dodoi{10.3847/1538-4357/ab6ffc}

% type= article
\bibitem[{L.~A. {Tarr} {et~al.}(2017){Tarr}, {Linton}, \&
  {Leake}}]{2017ApJ...837...94T}
{Tarr}, L.~A., {Linton}, M., \& {Leake}, J. 2017,
  \bibinfo{title}{{Magnetoacoustic Waves in a Stratified Atmosphere with a
  Magnetic Null Point},} \apj, 837, 94, \dodoi{10.3847/1538-4357/aa5e4e}

% type= article
\bibitem[{B.~J. {Thompson} {et~al.}(1998){Thompson}, {Plunkett}, {Gurman},
  {Newmark}, {St. Cyr}, \& {Michels}}]{1998GeoRL..25.2465T}
{Thompson}, B.~J., {Plunkett}, S.~P., {Gurman}, J.~B., {et~al.} 1998,
  \bibinfo{title}{{SOHO/EIT observations of an Earth-directed coronal mass
  ejection on May 12, 1997},} \grl, 25, 2465, \dodoi{10.1029/98GL50429}

% type= article
\bibitem[{J.~O. {Thurgood} \& J.~A. {McLaughlin}(2013){Thurgood} \&
  {McLaughlin}}]{2013A&A...555A..86T}
{Thurgood}, J.~O., \& {McLaughlin}, J.~A. 2013, \bibinfo{title}{{Nonlinear
  Alfv{\'e}n wave dynamics at a 2D magnetic null point: ponderomotive force},}
  \aap, 555, A86, \dodoi{10.1051/0004-6361/201321338}

% type= article
\bibitem[{Y. {Uchida}(1968){Uchida}}]{1968SoPh....4...30U}
{Uchida}, Y. 1968, \bibinfo{title}{{Propagation of Hydromagnetic Disturbances
  in the Solar Corona and Moreton's Wave Phenomenon},} \solphys, 4, 30,
  \dodoi{10.1007/BF00146996}

% type= article
\bibitem[{A.~M. {Veronig} {et~al.}(2010){Veronig}, {Muhr}, {Kienreich},
  {Temmer}, \& {Vr{\v{s}}nak}}]{2010ApJ...716L..57V}
{Veronig}, A.~M., {Muhr}, N., {Kienreich}, I.~W., {Temmer}, M., \&
  {Vr{\v{s}}nak}, B. 2010, \bibinfo{title}{{First Observations of a Dome-shaped
  Large-scale Coronal Extreme-ultraviolet Wave},} \apjl, 716, L57,
  \dodoi{10.1088/2041-8205/716/1/L57}

% type= article
\bibitem[{Y. {Voitenko} \& M. {Goossens}(2004){Voitenko} \&
  {Goossens}}]{2004ApJ...605L.149V}
{Voitenko}, Y., \& {Goossens}, M. 2004, \bibinfo{title}{{Cross-Field Heating of
  Coronal Ions by Low-Frequency Kinetic Alfv{\'e}n Waves},} \apjl, 605, L149,
  \dodoi{10.1086/420927}

% type= article
\bibitem[{T. {Wang} {et~al.}(2026){Wang}, {Liu}, {Ofman}, {Sun}, \&
  {Jin}}]{2026arXiv260313456W}
{Wang}, T., {Liu}, W., {Ofman}, L., {Sun}, X., \& {Jin}, M. 2026,
  \bibinfo{title}{{Quasi-Periodic Fast-Mode Wave Trains Associated with the
  2015-Jun-22 M6.5 Flare in AR\raisebox{-0.5ex}\textasciitilde12371:
  Observations and 3D MHD Modeling},} arXiv e-prints, arXiv:2603.13456,
  \dodoi{10.48550/arXiv.2603.13456}

% type= article
\bibitem[{A. {Warmuth}(2015){Warmuth}}]{2015LRSP...12....3W}
{Warmuth}, A. 2015, \bibinfo{title}{{Large-scale Globally Propagating Coronal
  Waves},} Living Reviews in Solar Physics, 12, 3, \dodoi{10.1007/lrsp-2015-3}

% type= article
\bibitem[{A. {Warmuth} \& G. {Mann}(2011){Warmuth} \&
  {Mann}}]{2011A&A...532A.151W}
{Warmuth}, A., \& {Mann}, G. 2011, \bibinfo{title}{{Kinematical evidence for
  physically different classes of large-scale coronal EUV waves},} \aap, 532,
  A151, \dodoi{10.1051/0004-6361/201116685}

% type= article
\bibitem[{D.~G. {Wentzel}(1974){Wentzel}}]{1974SoPh...39..129W}
{Wentzel}, D.~G. 1974, \bibinfo{title}{{Coronal Heating by Alfv{\'e}n Waves},}
  \solphys, 39, 129, \dodoi{10.1007/BF00154975}

% type= article
\bibitem[{G.~L. {Withbroe} \& R.~W. {Noyes}(1977){Withbroe} \&
  {Noyes}}]{1977ARA&A..15..363W}
{Withbroe}, G.~L., \& {Noyes}, R.~W. 1977, \bibinfo{title}{{Mass and energy
  flow in the solar chromosphere and corona.},} \araa, 15, 363,
  \dodoi{10.1146/annurev.aa.15.090177.002051}

% type= article
\bibitem[{A.~N. {Wright}(1991){Wright}}]{1991GeoRL..18.1951W}
{Wright}, A.~N. 1991, \bibinfo{title}{{MHD wave coupling in inhomogeneous
  media},} \grl, 18, 1951, \dodoi{10.1029/91GL02630}

% type= article
\bibitem[{C. {Xia} {et~al.}(2018){Xia}, {Teunissen}, {El Mellah}, {Chan{\'e}},
  \& {Keppens}}]{2018ApJS..234...30X}
{Xia}, C., {Teunissen}, J., {El Mellah}, I., {Chan{\'e}}, E., \& {Keppens}, R.
  2018, \bibinfo{title}{{MPI-AMRVAC 2.0 for Solar and Astrophysical
  Applications},} \apjs, 234, 30, \dodoi{10.3847/1538-4365/aaa6c8}

% type= article
\bibitem[{N. {Yadav} {et~al.}(2022){Yadav}, {Keppens}, \& {Popescu
  Braileanu}}]{2022A&A...660A..21Y}
{Yadav}, N., {Keppens}, R., \& {Popescu Braileanu}, B. 2022,
  \bibinfo{title}{{3D MHD wave propagation near a coronal null point: New wave
  mode decomposition approach},} \aap, 660, A21,
  \dodoi{10.1051/0004-6361/202142688}

% type= article
\bibitem[{T.~V. {Zaqarashvili}(2003){Zaqarashvili}}]{2003A&A...399L..15Z}
{Zaqarashvili}, T.~V. 2003, \bibinfo{title}{{Observation of coronal loop
  torsional oscillation},} \aap, 399, L15, \dodoi{10.1051/0004-6361:20030084}

% type= article
\bibitem[{T.~V. {Zaqarashvili} {et~al.}(2007){Zaqarashvili}, {Khutsishvili},
  {Kukhianidze}, \& {Ramishvili}}]{2007A&A...474..627Z}
{Zaqarashvili}, T.~V., {Khutsishvili}, E., {Kukhianidze}, V., \& {Ramishvili},
  G. 2007, \bibinfo{title}{{Doppler-shift oscillations in solar spicules},}
  \aap, 474, 627, \dodoi{10.1051/0004-6361:20077661}

% type= article
\bibitem[{T.~V. {Zaqarashvili} {et~al.}(2006){Zaqarashvili}, {Oliver}, \&
  {Ballester}}]{2006A&A...456L..13Z}
{Zaqarashvili}, T.~V., {Oliver}, R., \& {Ballester}, J.~L. 2006,
  \bibinfo{title}{{Spectral line width decrease in the solar corona: resonant
  energy conversion from Alfv{\'e}n to acoustic waves},} \aap, 456, L13,
  \dodoi{10.1051/0004-6361:20065805}

% type= article
\bibitem[{T.~V. {Zaqarashvili} \& B. {Roberts}(2002){Zaqarashvili} \&
  {Roberts}}]{2002PhRvE..66b6401Z}
{Zaqarashvili}, T.~V., \& {Roberts}, B. 2002, \bibinfo{title}{{Swing wave-wave
  interaction: Coupling between fast magnetosonic and Alfv{\'e}n waves},} \pre,
  66, 026401, \dodoi{10.1103/PhysRevE.66.026401}

% type= article
\bibitem[{T.~V. {Zaqarashvili} \& B. {Roberts}(2006){Zaqarashvili} \&
  {Roberts}}]{2006A&A...452.1053Z}
{Zaqarashvili}, T.~V., \& {Roberts}, B. 2006, \bibinfo{title}{{Two-wave
  interaction in ideal magnetohydrodynamics},} \aap, 452, 1053,
  \dodoi{10.1051/0004-6361:20053565}

% type= article
\bibitem[{Y. {Zhou} {et~al.}(2024){Zhou}, {Li}, \&
  {Keppens}}]{2024ApJ...968..123Z}
{Zhou}, Y., {Li}, X., \& {Keppens}, R. 2024, \bibinfo{title}{{Frozen-field
  Modeling of Coronal Condensations with MPI-AMRVAC. I. Demonstration in
  Two-dimensional Models},} \apj, 968, 123, \dodoi{10.3847/1538-4357/ad4466}

\end{thebibliography}
%\bibliographystyle{aasjournalv7}


\end{document}